\documentclass[reprint,twocolumn,showpacs,showkeys,preprintnumbers,amsmath,amssymb,floatfix,aps,pra,longbibliography]{revtex4-1}

\usepackage{graphicx}
\usepackage{bm}

\usepackage{physics}
\usepackage{placeins}
\usepackage{graphics}
\usepackage{color}
\usepackage[hidelinks]{hyperref}
\usepackage{multirow}
\usepackage{blindtext}
\usepackage{pdfpages}

\usepackage{xcolor}
\hypersetup{
    colorlinks,
    linkcolor={red!50!black},
    citecolor={blue!50!black},
    urlcolor={blue!80!black}
}

\makeatletter
\AtBeginDocument{\let\LS@rot\@undefined}
\makeatother  

\begin{document}


\title{Strong manipulation of the valley splitting upon twisting and gating \\ 
in MoSe$_2$/CrI$_3$ and WSe$_2$/CrI$_3$ van der Waals heterostructures}

\author{Klaus Zollner}
\email{klaus.zollner@physik.uni-regensburg.de}
\affiliation{Institute for Theoretical Physics, University of Regensburg, 93040 Regensburg, Germany}

\author{Paulo E. Faria~Junior}
\affiliation{Institute for Theoretical Physics, University of Regensburg, 93040 Regensburg, Germany}

\author{Jaroslav Fabian}
\affiliation{Institute for Theoretical Physics, University of Regensburg, 93040 Regensburg, Germany}

\begin{abstract}
Van der Waals (vdW) heterostructures provide a rich playground to engineer electronic, spin, and optical properties of individual two-dimensional materials. We investigate the twist-angle and gate dependence of the proximity-induced exchange coupling in the monolayer transition-metal dichalcogenides (TMDCs) MoSe$_2$ and WSe$_2$ due to the vdW coupling to the ferromagnetic semiconductor CrI$_3$, from first-principles calculations.
A model Hamiltonian, that captures the relevant band edges at the $K/K^{\prime}$ valleys of the proximitized TMDCs, is employed to quantify the proximity-induced exchange. Upon twisting from 0$^{\circ}$ to 30$^{\circ}$, we find a transition of the TMDC valence band (VB) edge exchange splitting from about $-2$ to $2$~meV, while the conduction band (CB) edge exchange splitting remains nearly unchanged at around $-3$~meV. For the VB of WSe$_2$ (MoSe$_2$) on CrI$_3$, the exchange coupling changes sign at around $8^{\circ}$ ($16^{\circ}$). We find that even at the angles with almost zero spin splittings of the VB, the real-space spin polarization profile of holes at the band edge is highly non-uniform, with alternating spin up and spin down orbitals. Furthermore, a giant tunability of the proximity-induced exchange coupling is provided by a transverse electric field of a few V/nm. 
Within our first-principles framework, we are limited to commensurate structures, 
and thus considered a maximum strain of 2.2\% for the individual monolayers.
By investigating different atomic stacking configurations of the strained supercells, we demonstrate that proximity exchange varies locally in space within experimentally realistic setups.
We complement our \textit{ab initio} results by calculating the excitonic valley splitting to provide experimentally verifiable optical signatures of the proximity exchange. Specifically, we predict that the valley splitting increases almost linearly as a function of the twist angle. Furthermore, the proximity exchange is highly tunable by gating, allowing to tailor the valley splitting in the range of 0 to 12~meV in WSe$_2$/CrI$_3$, which is equivalent to external magnetic fields of up to about 60~Tesla. Our results highlight the important impact of the twist angle and gating when employing magnetic vdW heterostructures in experimental geometries.
\end{abstract}

\maketitle

\section{Introduction}
Two-dimensional (2D) materials offer a huge variety of electronic, optical, spin, magnetic, and topological properties for future applications, which can be further tailored via external knobs\cite{Vincent2021:APR,Ma2021:AOM,Dieny2019:arxiv,Qiao2021:EEM,Sierra2021:NN,Roche2015:2DM,Schaibley2016:NRM,Wang2018:AFM,Briggs2019:2DM,Khan2020:JMCC,Wang2022:ACS}. 
For example, straining and gating of transition-metal dichalcogenides (TMDCs) can drive structural phase transitions, accompanied by significant modifications of the aforementioned properties \cite{Duerloo2014:NC,Song2016:NL,Hou2019:NN,Ye2012:S,Zhang2019:NM,Qian2014:S,Rehn2018:CM,Li2016:NC,Li2021:NRM}. Particularly,  Song \textit{et al.}~\cite{Song2016:NL} demonstrated a room temperature semiconductor-metal transition in thin film MoTe$_2$, due to the 2H-1T' structural phase transition via tensile strain of only 0.2\%. 
Furthermore, if one wants to preserve the electronic properties of a 2D monolayer, and on-demand engineer its band structure, the concept of van der Waals (vdW) heterostructures and proximity effects come into play \cite{Zutic2019:MT,Geim2013:Nat,Sierra2021:NN,Novoselov2016:Sci}. This concept is exemplified by graphene/TMDC heterostructures, in which the characteristic Dirac and semiconductor dispersions of the individual monolayers are well preserved, but the TMDC induces a giant (meV scale) proximity spin-orbit coupling (SOC) in graphene \cite{Gmitra2015:PRB,Gmitra2016:PRB}. In terms of applications, graphene/TMDC heterostructures can be employed for optospintronics \cite{Avsar2017:NL,Avsar2019:arxiv} and spin-charge conversion \cite{Offidani2017:PRL,Ghiasi2019:NL,Khokhriakov2020:NC,Herlin2020:APL}, which is only possible due to the weakly coupled bilayer system, nicely demonstrating the power of vdW engineering.

Monolayer TMDCs---MoS$_2$, WS$_2$, MoSe$_2$, and WSe$_2$---are interesting on their own, since they are air-stable 2D semiconductors, which have a direct band gap in the optical range at the $K/K^{\prime}$ valleys of the hexagonal Brillouin zone \cite{Kormanyos2014:2DM, Liu2015:CSR, Tonndorf2013:OE,Tongay2012:NL,Eda2011:NL}.
External stimuli can be employed to engineer the electronic structure and the optical transitions \cite{Chaves2020:2d,Ma2021:AOM}. For example, straining a monolayer WS$_2$ tunes the size of the band gap and leads to a direct-to-indirect band gap transition, as predicted by band structure calculations and observed in photoluminescence experiments \cite{Oliva2022:arxiv,Zollner2019:strain,Blundo2020:PRR,Blundo2021APR}.
Even though the optical excitations at $K/K^{\prime}$ valleys for unstrained monolayer TMDCs are energetically degenerate, they are helecity-sensitive --- due to the lack of inversion symmetry combined with the presence of time-reversal symmetry --- allowing to selectively address charge carriers at $K$ or $K^{\prime}$ \cite{Xiao2012:PRL}. The valley is thus another
potentially relevant parameter for information technology \cite{Vitale2018:S,Schaibley2016:NRM,Liu2019:NRF}. As experimentally demonstrated, valley polarizations of about 30\% can be achieved in monolayer MoS$_2$ by optical pumping \cite{Zeng2012:NN,Mak2012:NN}, enabling valleytronic devices \cite{Schaibley2016:NRM,Vitale2018:S,Bussolotti2018:NF,Langer2018:Nat,Liu2019:NRF}.

To achieve sizable valley polarizations, giant magnetic fields need to be applied. Indeed, a magnetic field of 1~Tesla yields only 0.1--0.2~meV of Zeeman splitting \cite{Srivastava2015:NP, Aivazian2015:NP, Li2014:PRL, MacNeill2015:PRL,Mitioglu2015:NL,Stier2016:NatComm,Wozniak2020:PRB,FariaJunior2022}.
Proximity exchange fields perform much better and can induce several meVs, without significantly altering the band structure of TMDCs 
\cite{Ye2016:NN, Zhong2017:SA, Ji2018:PCCP, Li2018:PCCP, Qi2015:PRB, Zhang2016:AM, Xu2018:PRB, Zhao2017:NN, Seyler2018:NL, Peng2017:ACS,Norden2019:NC,Zhao2017:NN,Ciorciaro2020:arxiv,Fang2022:PCCP}.
Of particular interest are all 2D vdW heterostructures, such as the recently considered WSe$_2$/CrI$_3$
bilayers \cite{Seyler2018:NL, Zhong2017:SA,Zhong2020:NN,Zhang2019:PRB,Zollner2019a:PRB,Dabrowski2022:NC}, which show a few meV of proximity exchange. Remarkably, recent experiments have demonstrated all-optical magnetization switching in such heterostructures~\cite{Dabrowski2022:NC}.
Moreover, an asymmetric magnetic proximity interaction has been recently observed in MoSe$_2$/CrBr$_3$ van der Waals heterostructures~\cite{Choi2022:NM}.
Nevertheless, most of these investigations have not considered the consequence of one important control parameter, which is the twist angle. 

Only recently the relative twist angle between the monolayers has been recognized as a crucial degree of freedom. The most prominent examples are twisted bilayer and trilayer graphene, exhibiting magnetism and superconductivity due to strong correlations \cite{Cao2018:Nat,Cao2018a:Nat,Arora2020:arxiv,Stepanov2020:Nat,Lu2019:Nat,Sharpe2019:SC, Saito2021:Nat,Serlin2020:S,Nimbalkar2020:NML,Bultinck2020:PRL,Repellin2020:PRL,Choi2019:NP,Lisi2021:NP,Balents2020:NP,Wolf2019:PRL,Zhu2020:PRL,Park2021:Nat,Chen2019:NP,Chen2020:Nat,Chen2019:Nat,Zhou2021:arxiv,Chou2021:arxiv,Phong2021:arxiv,Zhou2021:arxiv2,Qin2021:arxiv}. Other platforms for correlated physics are offered by twisted TMDCs \cite{Tang2020:Nat,Barman2022:ACS,Yu2021:PS} and twisted CrI$_3$ mono- and bilayers \cite{Xu2022:NN,Cheng2022:arxiv}. 
Regarding proximity effects, a recent study
shows the sensitivity of the spin polarization, magnetic anisotropy, and Dzyaloshinskii-Moriya interaction on the twist angle in graphene/2H-VSeTe heterostructures \cite{Yan2021:PE}. Similarly, theoretical \cite{David2019:arxiv,Li2019:PRB,Naimer2021:arxiv,Pezo2021:arXiv,Lee2022:arxiv,Peterfalvi2021:arxiv,Veneri2022:arxiv} and experimental \cite{Ernandes2021:NM} studies show that the strength of proximity SOC in graphene/TMDC and the proximity exchange in graphene/Cr$_2$Ge$_2$Te$_6$ \cite{Zollner2022:PRL} heterostructures can be tuned by the twist angle.

In our previous work  \cite{Zollner2019a:PRB} on TMDC/CrI$_3$ bilayers we have only considered two twist angles, 0$^{\circ}$ and 30$^{\circ}$, predicting a significant (meV scale) proximity exchange, with different sign for the TMDC valence band (VB) edge spin splitting at the two twist angles. No detailed study of the effect has been done for intermediate twist angles to see if the sign change is systematic or accidental. Another important question relates to the proximity exchange of the conduction band (CB): does the direction of the proximity exchange remain independent of the twist angle, as suggested by the 0$^{\circ}$ and 30$^{\circ}$ predictions? What is the variation of the magnitude of the proximity exchange as a function of the twist angle? Could there be a scenario such as predicted for
graphene/Cr$_2$Ge$_2$Te$_6$ \cite{Zollner2022:PRL} that the effective proximity exchange becomes antiferromagnetic at an intermediate angle? What are the real-space spin-polarization profiles of the conduction and valence electrons in the TMDCs for different twist angles? In the present study, we systematically address these questions.

Recent band structure investigations of WSe$_2$/CrI$_3$ heterostructures by Ge \textit{et al.} \cite{Ge2022:CM} revealed a relative valley splitting enhancement of about 1000\% by twisting, in good agreement with Ref.~\cite{Zollner2019a:PRB}. They have shown that mainly the TMDC VB splitting is responsible for the massive twist-tunability of the valley splitting. In addition they also found a reversal of the VB splitting and related valley splitting upon twisting, but did not analyze this fact.  Overall, the tunability of the valley splitting is attributed to the magnetic proximity effect and related to the enhancement of the Cr magnetic moments upon twisting. 
However, Ge \textit{et al.} \cite{Ge2022:CM} considered just 4 twist angles and rather large strains of up to 5.5\%, with respect to the lattice constants determined from density functional theory (DFT), which significantly alters both monolayer dispersions and CrI$_3$ magnetic properties \cite{Zollner2019:strain,Wu2019:PCCP}. In addition, an electric field dependence was performed only for one specific twist angle of 16.1$^{\circ}$.

Here, we systematically investigate the twist-angle and electric-field dependence of the proximity-induced exchange coupling in MoSe$_2$/CrI$_3$ and WSe$_2$/CrI$_3$ bilayers, from first principles. 
We consider 7 different angles between 0$^{\circ}$ and 30$^{\circ}$ with a maximum strain---needed for commensurability of the unit cells---of about 2.2\% applied to the monolayers when employing experimentally determined lattice constants. 
Within the heterostructures the TMDCs preserve their characteristic dispersion, while the magnetic insulator substrate provides proximity exchange, splitting the relevant TMDC VB and CB edges (in the absence of SOC, see below) at $K/K^{\prime}$ valleys. Depending on the relative twist angle the band edge splittings can be markedly different, and even the spin ordering can be reversed. In particular, when twisting from 0$^{\circ}$ to 30$^{\circ}$, the TMDC CB edge splitting remains nearly unchanged at around $-3$~meV. In contrast, there is a smooth transition of the TMDC VB edge spin splitting from about $-2$ to 2~meV (the spin ordering changes). The finding about the reversal of the spin ordering is consistent with Ref.~\cite{Zollner2019a:PRB}, but here we additionally find that it happens in an almost linear fashion upon twisting, and we identify the twist angles at which the transition happens: about 8$^{\circ}$ (16$^{\circ}$) for WSe$_2$ (MoSe$_2$).
The origin of this reversal is traced back to the twist-angle dependent backfolding of the TMDC $K/K^{\prime}$ valleys into the CrI$_3$ Brillouin zone.
By investigating different atomic stacking configurations of the strained supercells, we show that proximity exchange varies locally in space within experimentally realistic heterostructures.

Furthermore, we reveal a rather high tunability of the proximity-induced exchange coupling by applying a transverse electric field of a few V/nm across all the twisted WSe$_2$/CrI$_3$ heterostructures. In Ref.~\cite{Zollner2019a:PRB}, the electric field tunability was investigated for the 0$^{\circ}$ structure only.
We also introduce a minimal model Hamiltonian to describe the twist- and gate-tunable proximity effects in the TMDCs due to CrI$_3$. The Hamiltonian together with the fitted parameters provide an effective description for the TMDC band edge physics at $K/K^{\prime}$ valleys, which is relevant for calculating magneto-optical Kerr effect~\cite{Catarina2020:2DM,Henriques2020:PRB}, absorption spectra~\cite{Scharf2017:PRL,Zollner2019a:PRB,Zollner2020:PRB}, or coupled spin Hall and valley Hall effects~\cite{Xiao2012:PRL}.
Finally, we give specific predictions for experimentally verifiable optical signatures of the proximity exchange effects, by calculating the excitonic absorption spectra employing the Bethe-Salpeter equation. In addition to 0$^{\circ}$ and
30$^{\circ}$ angles~\cite{Zollner2019a:PRB}, we provide the full twist-angle and electric-field dependence of the valley splitting. We find a rather high tunability of the valley splitting of the first intralayer exciton peak, ranging from 0 to 12~meV by gating and twisting in WSe$_2$/CrI$_3$ bilayers, equivalent to external magnetic fields of up to about 60~Tesla.

    \begin{figure}[htb]
     \includegraphics[width=.9\columnwidth]{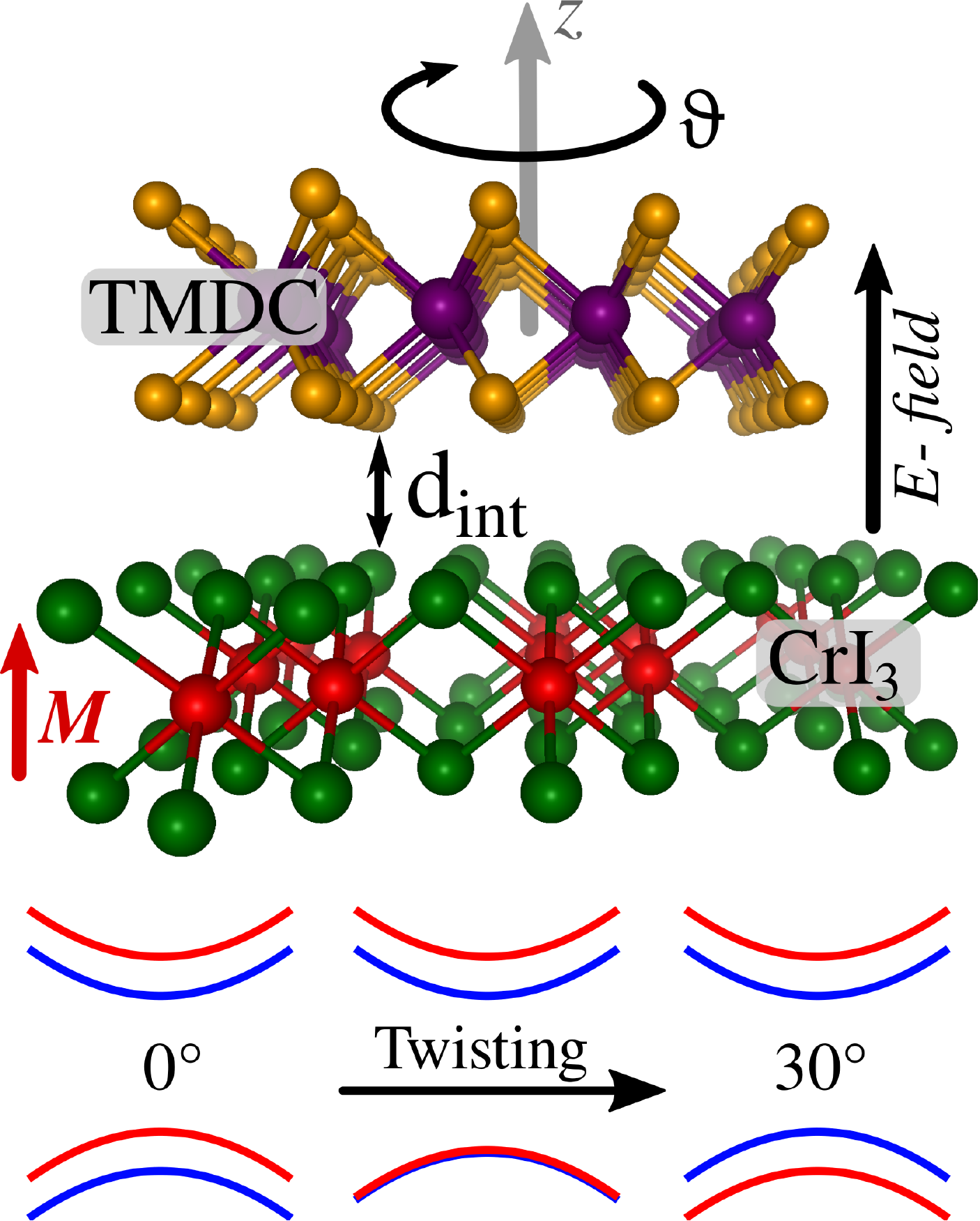}
     \caption{3D view of a TMDC (MoSe$_2$ or WSe$_2$) on CrI$_3$, where we define the interlayer distance, $\mathrm{d}_{\mathrm{int}}$. We twist the TMDC by an angle $\vartheta$ around the $z$ axis, with respect to the magnetic semiconductor CrI$_3$, with magnetization $\bm{M}$ along the $z$ direction.
     The twist-angle evolution of the proximitized TMDC band edges are sketched. Red bands are polarized spin up, while blue bands are polarized spin down.
     By twisting from 0$^{\circ}$ to 30$^{\circ}$, the TMDC VB edge splitting, which is due to proximity exchange, first vanishes and then reverses sign. 
    }\label{Fig:Structure}
    \end{figure}

The paper is organized as follows. In Sec.~\ref{sec:geometry_comp}, we first address the structural setup and summarize the calculation details for obtaining the electronic structures of the twisted TMDC/CrI$_3$ bilayers. In Sec.~\ref{sec:model}, we introduce the model Hamiltonian that captures the relevant band edges at $K/K^{\prime}$ valleys of the proximitized TMDCs, which are used to fit the first-principles results. In Sec.~\ref{sec:proximity_exch}, we show and discuss exemplary calculated electronic structures, along with the model Hamiltonian fits. The twist-angle and gate dependence of the proximity-induced exchange couplings is also addressed. In Sec.~\ref{sec:valley_splitting}, we show and discuss the twist-angle and electric-field tunability of the valley splitting, as calculated from the absorption spectra employing the Bethe-Salpeter equation. 
Finally, in Sec.~\ref{sec:summary} we conclude the manuscript.

\section{Geometry Setup \& Computational Details}
\label{sec:geometry_comp}

    \begin{table*}[htb]
    \caption{\label{Tab:crystal} Structural information for the TMDC/CrI$_3$ heterostructures. 
    We list the twist angle $\vartheta$ between the layers, the number of atoms (NoA) in the heterostructure supercell, the number $n_k$ for the $k$-point sampling, the lattice constants and biaxial strains $\varepsilon$ applied to the TMDCs and CrI$_3$, the calculated dipoles and the relaxed interlayer distances $d_{\textrm{int}}$ of the MoSe$_2$ (WSe$_2$) structures.}
    \begin{ruledtabular}
    \begin{tabular}{ccccccccc}
     $\vartheta$ [°] & NoA & $n_k$ &  a$_\textrm{TMDC}$ [\AA] & $\varepsilon_\textrm{TMDC}$ [\%]  & a$_\textrm{CrI3}$ [\AA] & $\varepsilon_\textrm{CrI3}$ [\%] & dipole [debye] & $d_{\textrm{int}}$ [\AA] \\ \hline
     0.0000 & 20 & 30  & 3.3608 & 2.21 & 6.7215 & -2.12 & 0.0507 (0.1082) & 3.5467 (3.5000) \\
     8.2132 & 243 & 9 & 3.3400 & 1.58 & 6.7472 & -1.75 & 0.0322 (0.5475) & 3.5888 (3.5661) \\
     10.1583 & 149 & 12 & 3.2880 & 0 & 6.9189 &	0.76 & 0.3590 (0.6993) & 3.5858 (3.5749) \\
     16.1021 & 63 & 24 & 3.2880 & 0 & 6.8444 & -0.33 & 0.0881 (0.2474) & 3.5930 (3.5490) \\
     21.7868 & 140 & 12 & 3.3610 & 2.22 & 6.7189 & -2.16 & -0.0725 (0.2432) & 3.5666 (3.5499) \\
     25.2850 & 183 & 9 & 3.3350 & 1.43 & 6.7604 & -1.55 & 0.0033 (0.4402) & 3.5885 (3.5627)\\
     30.0000 & 243 & 9 & 3.3400 & 1.58 & 6.7492 & -1.72 & -0.0347 (0.5360) & 3.5837 (3.5640)\\
    \end{tabular}
    \end{ruledtabular}
    \end{table*}

    The twisted MoSe$_2$/CrI$_3$ and WSe$_2$/CrI$_3$ heterostructures for the DFT calculations are set-up with the {\tt atomic simulation environment (ASE)} \cite{ASE} and the {\tt CellMatch} code \cite{Lazic2015:CPC}, implementing the coincidence lattice method \cite{Koda2016:JPCC,Carr2020:NRM}.
    Monolayers of TMDCs and CrI$_3$ are based on hexagonal unit cells, with lattice constants of $a = 3.288$~\AA~ (MoSe$_2$), $a = 3.282$~\AA~ (WSe$_2$), and $a = 6.867$~\AA~ (CrI$_3$) 
    \cite{McGuire2015:CM, Schutte1987:JSSC, James1963:AC}, which need to be strained in the twisted heterostructures, in order to form commensurate supercells for periodic DFT calculations. Since MoSe$_2$ and WSe$_2$ have nearly the same lattice constant, we set them as equal in the following. 
    In Table~\ref{Tab:crystal} we summarize the main structural information for the twist angles we consider. All our heterostructures have $C_3$ symmetry. In Fig.~S1~\footnotemark[1], we show top views of all employed supercell geometries.
    In total, we investigate 7 different angles between 0$^{\circ}$ and 30$^{\circ}$. Especially these angles are suitable for DFT calculations, since strain applied to the monolayers is at maximum 2.2\%. 
    We already know that biaxial strain strongly influences the band gap, spin-orbit splittings and spin-valley properties of monolayer TMDCs \cite{Zollner2019:strain,FariaJunior2022}, as well as the band gap of CrI$_3$ \cite{Wu2019:PCCP}, therefore we keep the strain as small as possible.
    In addition, the number of atoms is kept below 250. Otherwise, also
    other angles could be investigated, but beyond reasonable strain limits and above a computationally feasible number of atoms in the structure.

	The electronic structure calculations and structural relaxation of the TMDC/CrI$_3$ heterostructures
	are performed by DFT~\cite{Hohenberg1964:PRB} 
	with {\tt Quantum ESPRESSO}~\cite{Giannozzi2009:JPCM,Giannozzi2017:JPCM}.
	Self-consistent calculations are carried out with a $k$-point sampling of 
	$n_k\times n_k\times 1$. The number $n_k$ is listed in Table~\ref{Tab:crystal} for all twist angles and depends on the number of atoms in the heterostructure.
	In addition, $n_k$ is limited by our computational power. Nevertheless, for large supercells the heterostructure Brillouin Zone is small and only a few $k$-points are necessary to get converged results. 
	
	We perform open shell calculations that provide the
	spin-polarized ground state of the CrI$_3$ monolayer. 
	A Hubbard parameter of $U = 3.0$~eV is used for Cr $d$-orbitals \cite{Zollner2019a:PRB,Jiang2021:APR}.
	We use an energy cutoff for charge density of $520$~Ry and
	the kinetic energy cutoff for wavefunctions is $65$~Ry for the scalar relativistic pseudopotentials
	with the projector augmented wave method~\cite{Kresse1999:PRB} with the 
	Perdew-Burke-Ernzerhof exchange correlation functional~\cite{Perdew1996:PRL}.
	SOC is neglected, since we are mainly interested in the twist-angle dependence of the proximity-induced exchange coupling. Moreover, we have already demonstrated that SOC and proximity exchange are rather independent in the investigated bilayers \cite{Zollner2019a:PRB}.
	For the relaxation of the heterostructures, we add DFT-D2 vdW corrections~\cite{Grimme2006:JCC,Grimme2010:JCP,Barone2009:JCC} and use 
	quasi-Newton algorithm based on trust radius procedure. 
	Dipole corrections \cite{Bengtsson1999:PRB} are also included to get correct band offsets and internal electric fields.
	In order to simulate quasi-2D systems, we add a vacuum of about $24$~\AA~to avoid interactions between periodic images in our slab geometry. For proper interlayer distances, we allow the atoms of the TMDCs, as well as the Cr atoms of the CrI$_3$, 
    to relax their $z$ coordinates, while
    the I atoms are allowed to move freely, because they form a distorted octahedral surrounding around the Cr atoms \cite{Jiang2018:NL}.  Relaxation is performed until every component of each force is reduced below $10^{-3}$~[Ry/$a_0$], where $a_0$ is the Bohr radius.
	
	After relaxation, we calculate the mean interlayer distances, $d_{\textrm{int}}$, from the $z$ coordinates of interfacial Se and I atoms. The interlayer distances are nearly independent of the twist angle, about $3.55$~\AA, and are listed in Table~\ref{Tab:crystal}.
	In Fig.~\ref{Fig:Structure}, we show the general structural setup of our TMDC/CrI$_3$ heterostructures, where the TMDC resides above CrI$_3$, with the magnetization $\bm{M}$ along $z$ direction specifying the spin quantization axis (spin up = $z$, spin down = $-z$).
    When we apply the transverse electric field (modeled by a zigzag potential), a positive field also points along $z$ direction.

\section{Model Hamiltonian}
\label{sec:model}
We want to describe proximity exchange effects in the monolayer TMDCs that are due to the magnetic insulator substrate CrI$_3$.
Similar to our recent work \cite{Zollner2019a:PRB}, we employ a minimal model Hamiltonian to describe 
the band edges of the TMDC close to $K$ and $K^{\prime}$ valleys, in the presence of proximity exchange
\begin{flalign}
\label{Eq:Hamiltonian}
&\mathcal{H} = \mathcal{H}_{0}+\mathcal{H}_{\Delta}+\mathcal{H}_{\textrm{ex}}\\
&\mathcal{H}_{0} = \hbar v_{\textrm{F}} s_0 \otimes (\tau\sigma_{x}k_{x}+\sigma_{y}k_{y}),\\
&\mathcal{H}_{\Delta} = \frac{\Delta}{2}s_{0}\otimes \sigma_{z},\\
&\mathcal{H}_{\textrm{ex}} = -s_{z} \otimes (B_{\textrm{c}}\sigma_{+} + B_{\textrm{v}} \sigma_{-}).
\end{flalign}
The valley index is $\tau = \pm 1$ for $K$/$K^{\prime}$ point and $v_{\textrm{F}}$ is the Fermi velocity. 
The Cartesian components $k_{x}$ and $k_{y}$ of the electron wave vector are measured from $K$/$K^{\prime}$. 
The pseudospin Pauli matrices are $\sigma_{\textrm{i}}$ acting on
the [conduction band (CB), valence band (VB)] subspace and spin Pauli matrices are $s_{\textrm{i}}$ acting on the ($\uparrow, \downarrow$) subspace,
with $\textrm{i} = \{0,x,y,z\}$.
The parameter $\Delta$ denotes the orbital gap of the spectrum. 
For short notation we introduce $ \sigma_{\pm} = \frac{1}{2}(\sigma_{0} \pm \sigma_{z})$. 
With a magnetic substrate layer, proximity exchange coupling will split the TMDC CB edge by $2B_{\textrm{c}}$ and the VB edge by $2B_{\textrm{v}}$.
The four basis states we use are $|\Psi_{\textrm{CB}}, \uparrow\rangle$, 
$|\Psi_{\textrm{VB}}^{\tau}, \uparrow\rangle$, $|\Psi_{\textrm{CB}}, \downarrow\rangle$, 
and $|\Psi_{\textrm{VB}}^{\tau}, \downarrow\rangle$. The wave functions are 
$|\Psi_{\textrm{CB}}\rangle = |d_{z^2}\rangle$ and 
$|\Psi_{\textrm{VB}}^{\tau}\rangle = \frac{1}{\sqrt{2}}(|d_{x^2-y^2}\rangle+\textrm{i}\tau |d_{xy}\rangle)$, 
corresponding to CB and VB at $K$/$K^{\prime}$ \cite{Kormanyos2014:2DM}.

\section{Twist- and gate-tunable proximity exchange}	
\label{sec:proximity_exch}
\subsection{Twist angle dependence}

\begin{figure*}[!htb]
 \includegraphics[width=.99\textwidth]{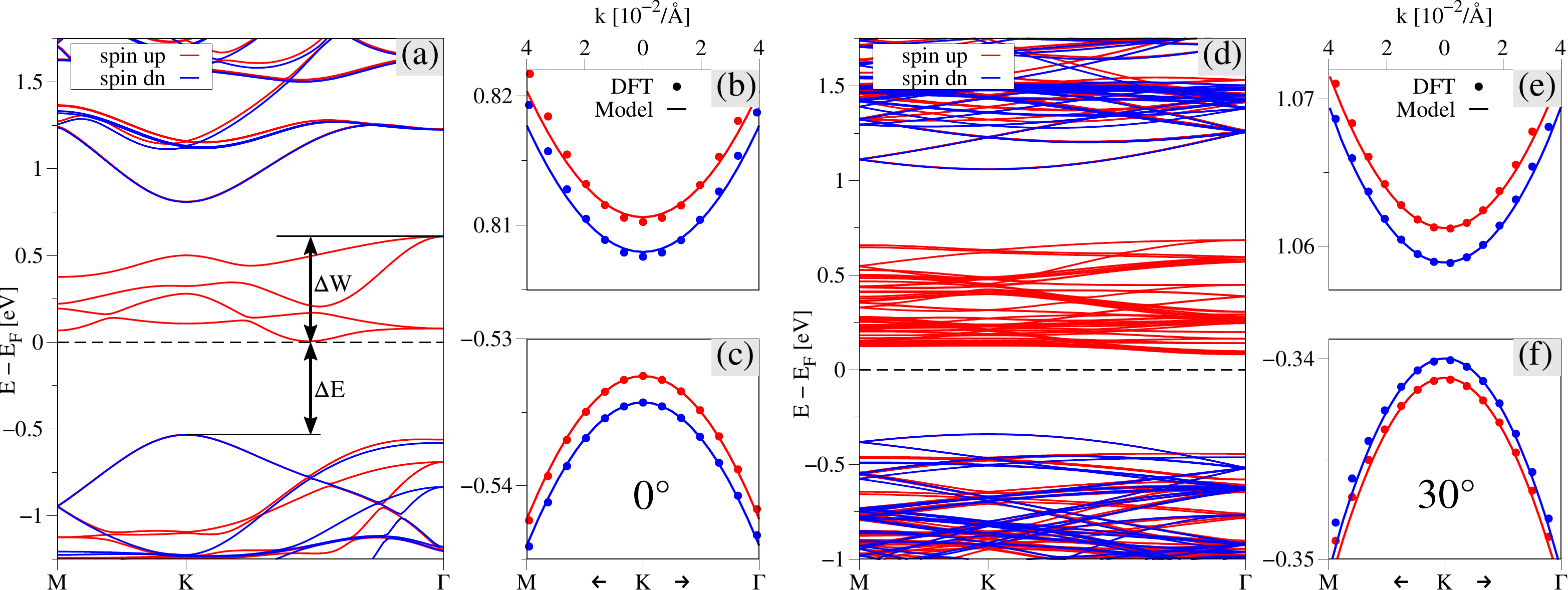}
 \caption{(a) DFT-calculated band structure of the MoSe$_2$/CrI$_3$ heterostructure for a twist angle of 0$^{\circ}$ along the high-symmetry path $M-K-\Gamma$.
 Bands in red are spin up, while bands in blue are spin down.
 We define the heterostructure band gap $\Delta E$, between the TMDC VB edge at K, and the minimum of spin polarized CrI$_3$ in-gap states. We also define the bandwidth $\Delta W$ of the in-gap states. (b) Zoom to the MoSe$_2$ CB edge near the $K$ point, showing proximity exchange split bands. 
 Symbols are DFT data and solid lines are the fitted model Hamiltonian results. (c) Same as (b), but for the VB edge. [(d)-(f)] Same as [(a)-(c)], but for a twist angle of 30$^{\circ}$. Comparing (c) and (f), the spin ordering of bands is reversed.
 }\label{Fig:bands_MoSe2_CrI3}
\end{figure*}

    \begin{table*}[htb]
    \caption{\label{Tab:fit} Fit parameters of Hamiltonian $\mathcal{H}$ for the TMDC/CrI$_3$ heterostructures 
for different twist angles. We summarize the Fermi velocity $v_{\textrm{F}}$, the orbital gap $\Delta$, proximity exchange parameters $B_{\textrm{c}}$ and $B_{\textrm{v}}$, the heterostructure band gap $\Delta E$, and the CrI$_3$ in-gap states bandwidth $\Delta W$, as defined in Fig.~\ref{Fig:bands_MoSe2_CrI3}(a).}
    \begin{ruledtabular}
    \begin{tabular}{cccccccc}
    & $\vartheta$ [°] & $v_{\textrm{F}}$ [$10^5$ m/s] & $\Delta$ [eV] & $B_{\textrm{c}}$ [meV] & $B_{\textrm{v}}$ [meV] & $\Delta E$ [eV] & $\Delta W$ [eV] \\
    \hline
MoSe$_2$ & 0.0000 & 4.358 & 1.343 & -1.343 & -0.901 & 0.538 & 0.603 \\
& 8.2132 & 4.631 & 1.403 & -1.070 & -0.262 & 0.432 & 0.601 \\
& 10.1583 &  4.626 & 1.555 & -1.427 & -0.365 & 0.330 & 0.485 \\
& 16.1021 &  4.622 & 1.556 & -1.332 & -0.012 & 0.337 & 0.535 \\
& 21.7868 &  4.575 & 1.340 & -1.253 & 0.213 & 0.463 & 0.617 \\
& 25.2850 & 4.048 & 1.415 & -1.221 & 0.415 & 0.417 & 0.587 \\
& 30.0000 & 4.622 & 1.401 & -1.176 & 0.484 & 0.422 & 0.603\\ 
\hline
WSe$_2$ & 0.0000 & 5.873 & 1.403 & -1.460 & -0.918 & 0.377 & 0.600 \\
& 8.2132 & 5.954 & 1.479 & -1.175 & -0.020 & 0.255 & 0.601 \\
& 10.1583 & 6.093 & 1.674 & -1.553 & 0.099 & 0.165 & 0.484 \\
& 16.1021 &  6.084 & 1.674 & -1.500 & 0.705 & 0.178 & 0.535 \\
& 21.7868 &  5.839 & 1.403 & -1.318 & 0.894 & 0.291 & 0.615 \\
& 25.2850 & 5.216 & 1.495 & -1.290 & 1.415 & 0.248 & 0.587 \\
& 30.0000 & 5.940 & 1.477 & -1.226 & 1.525 & 0.251 & 0.604 \\
    \end{tabular}
    \end{ruledtabular}
    \end{table*}

In Fig.~\ref{Fig:bands_MoSe2_CrI3}(a), we show the DFT-calculated band structure of the MoSe$_2$/CrI$_3$ heterostructure for a twist angle of 0$^{\circ}$. 
The band edges of the TMDC can be nicely recognized at the $K$ point at about $-0.5$ and $0.8$~eV with respect to the Fermi level. The states ranging from about $0$ to $0.5$~eV are spin polarized in-gap states that originate from the CrI$_3$ layer. 
Therefore, the heterostructure forms a type-II band alignment with a bandgap $\Delta E$ of about $0.5$~eV. 
In our heterostructures, small charge transfer is present between the monolayers, leading to a finite intrinsic electric (dipole) field. 
The dipole of the heterostructure, which controls the band alignment and $\Delta E$, depends on the twist angle, as summarized in Table~\ref{Tab:fit}. However, we believe that the strain that is applied to the monolayers has the most significant impact on the band alignment, as previous studies of graphene-based heterostructures suggest \cite{Zollner2022:PRL,Naimer2021:arxiv}. Therefore, the intrinsic band alignment that should be present in experiments can be read off from the 16.1$^{\circ}$ structure, with marginally strained monolayers (0.33\% of strain). 
In addition, the CrI$_3$ in-gap states bandwidth $\Delta W$ depends on the lattice constant used in the heterostructure. Compressive (tensile) strain increases (decreases) $\Delta W$.
Finally, also the TMDC band gap $\Delta$ depends on the lattice constant and is strain tunable \cite{Zollner2019:strain}. However, the 0$^{\circ}$ and 30$^{\circ}$ structures have rather similar strain, and still the proximity exchange is opposite for the VB; this strongly suggests that the variation of the proximity exchange with the twist angle is not much affected by the modest strain we have in the supercells.

\begin{figure}[!htb]
 \includegraphics[width=.9\columnwidth]{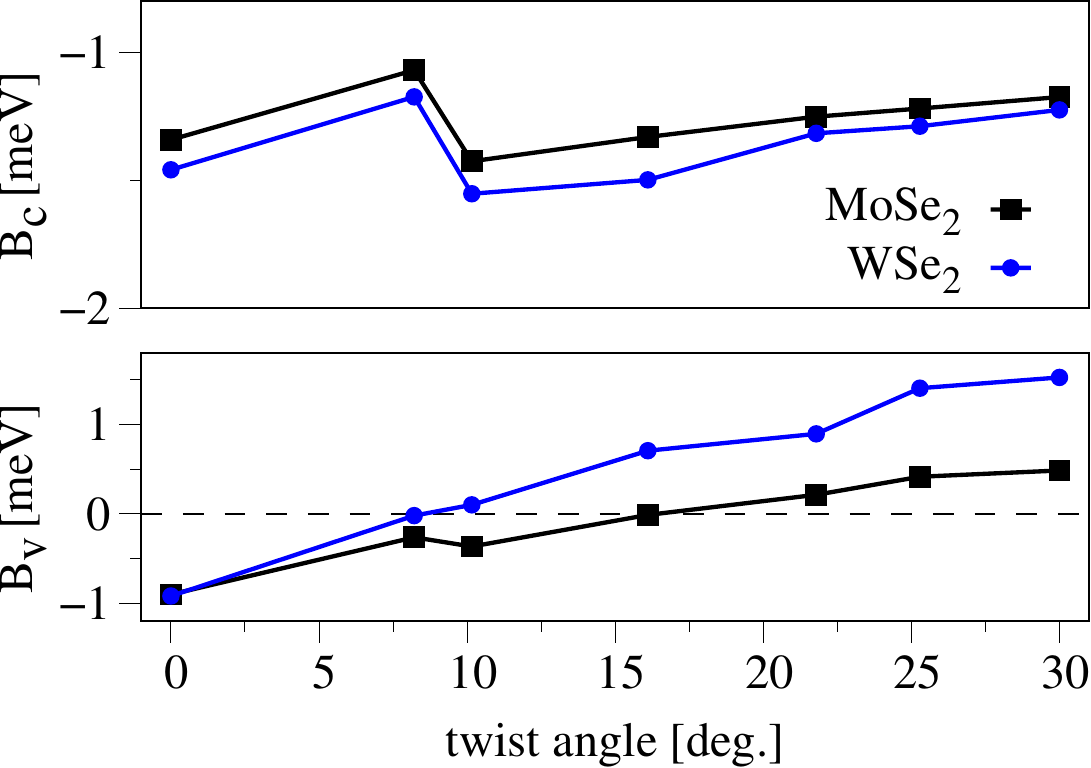}
 \caption{Calculated twist-angle dependence of the proximity exchange couplings $B_{\textrm{c}}$ and $B_{\textrm{v}}$ for the MoSe$_2$/CrI$_3$ and WSe$_2$/CrI$_3$ heterostructures. 
 }\label{Fig:BC_BV_vs_angle}
\end{figure}

In Figs.~\ref{Fig:bands_MoSe2_CrI3}[(b) and (c)], we show the zooms to the MoSe$_2$ band edges near the $K$ point for the 0$^{\circ}$ structure. The bands experience proximity-induced exchange splitting of about $2$~meV. The spin ordering is the same for CB and VB, with the spin down band lower in energy. The fit parameters from Table~\ref{Tab:fit} nicely reproduce the TMDC band edges. 
In Figs.~\ref{Fig:bands_MoSe2_CrI3}[(d)-f)], we summarize the results for the MoSe$_2$/CrI$_3$ heterostructure for a twist angle of 30$^{\circ}$. 
Overall, the band structure features remain the same, but with many more bands due to the larger supercell size compared to the 0$^{\circ}$ structure. 
Looking at the relevant TMDC band edges, proximity exchange splitting is still present, but the VB spin ordering is \textit{reversed} compared to the 0$^{\circ}$ case. The CB proximity exchange remains the same. Below and in the Supplemental Material (SM) \footnotemark[1] we further elaborate on the origin of this reversal.

In Fig.~\ref{Fig:BC_BV_vs_angle}, we summarize the twist-angle dependence of the proximity exchange parameters, $B_{\textrm{c}}$ and $B_{\textrm{v}}$, as listed in Table~\ref{Tab:fit} for the MoSe$_2$ and WSe$_2$ heterostructures. 
We find that the twist angle barely influences the CB proximity exchange parameter $B_{\textrm{c}}$, being fixed at around $-1.2$~meV. In contrast, the VB proximity exchange parameter, $B_{\textrm{v}}$, is negative for 0$^{\circ}$, \textit{vanishes and reverses sign} at about 16$^{\circ}$ (8$^{\circ}$) for MoSe$_2$ (WSe$_2$), and is positive for a twist angle of 30$^{\circ}$. The atomic arrangement (stacking) of the monolayers, for a fixed twist angle, can also influence proximity exchange \cite{Zollner2019a:PRB,Zhang2019:PRB}, see Supplemental Material~\footnotemark[1].
Apparently, for the different twist angles, we have chosen the stackings such that $B_{\textrm{c}}$ shows a kink around 8$^{\circ}$.

\begin{figure}[htb]
     \includegraphics[width=.99\columnwidth]{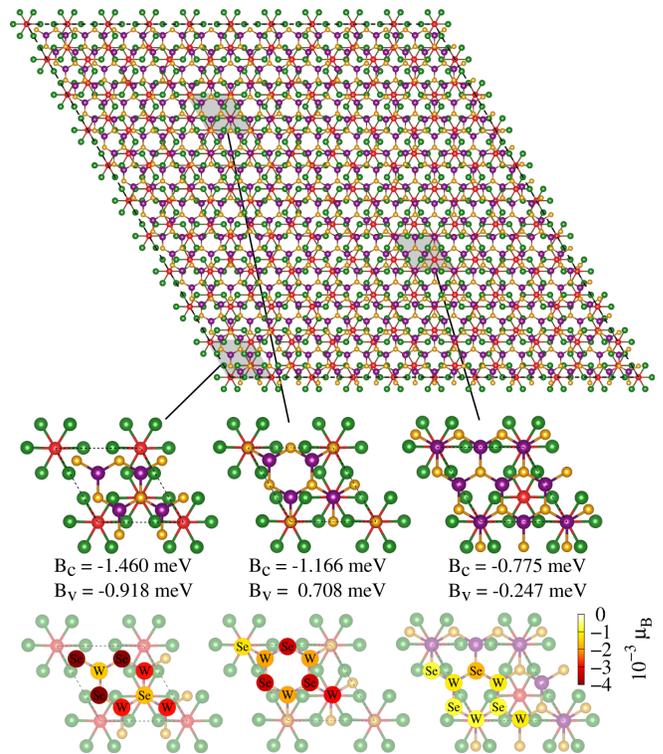}
     \caption{Top: An experimentally realistic $0^{\circ}$ WSe$_2$/CrI$_3$ heterostructure (2123 atoms), where we consider a $21\times 21$ supercell of WSe$_2$ on a $10\times 10$ supercell of CrI$_3$, corresponding to strains of 0\% and 0.5\%. Middle: Different high-symmetry stackings of the $0^{\circ}$ strained heterostructure are locally indicated, with their corresponding proximity exchange parameters $B_{\textrm{c}}$ and $B_{\textrm{v}}$. Bottom: Calculated proximity-induced local magnetic moments (colored spheres) of the interfacial Se and W atoms for the different stackings.
    }\label{Fig:Local_stackings}
    \end{figure}

For completeness, in the SM \footnotemark[1] we investigate different atomic stacking configurations and also briefly address the twist angles between 30$^{\circ}$ to 60$^{\circ}$. Overall, we find that in large supercells the stacking only marginally influences proximity exchange couplings, while in small supercells --- such as the 0$^{\circ}$ structure with only 20 atoms --- the VB edge splitting can be tuned in sign and magnitude by the atomic arrangement, see also Fig.~\ref{Fig:Local_stackings}. 
We believe that an averaging effect takes place in larger supercell geometries, where a lot of atoms are involved in forming the TMDC band edge states, which can locally pick up different exchange fields, but globally lead to the same band edge splitting. In other words, experimentally the proximity exchange will vary locally in space across the heterostructure. 

Such a local variation can be seen in Fig.~\ref{Fig:Local_stackings} (Fig.~S2), where we also show the proximity-induced local magnetic moments of interfacial Se and W atoms of different 0$^{\circ}$ stacking configurations (of all investigated heterostructures).
We find that the induced magnetic moments depend sensitively on the stacking and correlate with the proximity exchange parameters $B_{\textrm{c}}$ and $B_{\textrm{v}}$. 
With high-resolution vector magnetometry and gradiometry, based on nitrogen-vacancy centers in diamond \cite{Balasubramanian2008:N, Wang2015:NC2,Huxter2022:arxiv}, one can potentially resolve the demonstrated local variations of proximity exchange.

In Fig.~\ref{Fig:Local_stackings} and Fig.~S19, we explicitly map different high-symmetry stackings of our strained 0$^{\circ}$ structure to an experimentally more realistic setup, further confirming the local variation of proximity exchange.
In fact, the proximity exchange parameter $B_{\textrm{c}}$ varies by a factor of 2 among the investigated stackings, while the parameter $B_{\textrm{v}}$ can be even reversed in sign and magnitude. 
However, one has to relate this with experimental limitations. 
Even though there can be local variations of proximity exchange, typical photoluminescence techniques are not capable of addressing only a specific stacking. Then, even for the 0$^{\circ}$ case, an ensemble of stackings will be probed, thus leading to an average proximity exchange. Based on the findings for different stackings, the average exchange splittings follow the same predictions as we provide in Fig.~\ref{Fig:BC_BV_vs_angle} (VB and CB splittings negative for 0$^{\circ}$).
Moreover, twisting from $30^{\circ}$ to $60^{\circ}$ leads to similar proximity exchange couplings --- apart from differences in magnitude owed to the atomic stacking --- as twisting from $30^{\circ}$ to $0^{\circ}$.

As mentioned above, the strain is a byproduct of our structural setup, leading to nontrivial influences on band gaps and magnetic properties of our monolayers \cite{Zollner2019:strain, Wu2019:PCCP}. In the SM \footnotemark[1], we also investigate the influence of the strain distribution for the WSe$_2$/CrI$_3$ structure with $0^{\circ}$ twist angle. We find that the exchange parameters can be changed by at maximum $\pm 20$\% by tuning the strain, but their signs and the order of magnitude remains. In other words, the presented results are representative for experimental conditions (no or weak strain). Based on the findings for $0^{\circ}$, one can expect a similar strain dependence for the other investigated twist angles, where strain is similar or smaller compared to $0^{\circ}$, see Table~\ref{Tab:crystal}.
Especially for the $10.2^{\circ}$ and $16.1^{\circ}$ structures, where strain is small, we believe our calculations provide a close correspondence to experimental results. 

Regarding experiments, one can expect similar results for few layer CrI$_3$, including even and odd number of layers. In Ref.~\cite{Zollner2019a:PRB}, we have shown that the proximity exchange coupling is restricted to interfacial layers, i.~e., mainly the topmost CrI$_3$ layer influences the TMDC. In particular, we have shown that proximity exchange for TMDC on bilayer CrI$_3$ in ferro- and antiferromagnetic configuration is essentially unchanged. Of course, the sign of the proximity effect does depend on the magnetization direction of the interfacial CrI$_3$ layer, but not the magnitude.

\subsection{Electric field tunability}	

    \begin{figure}[htb]
     \includegraphics[width=.99\columnwidth]{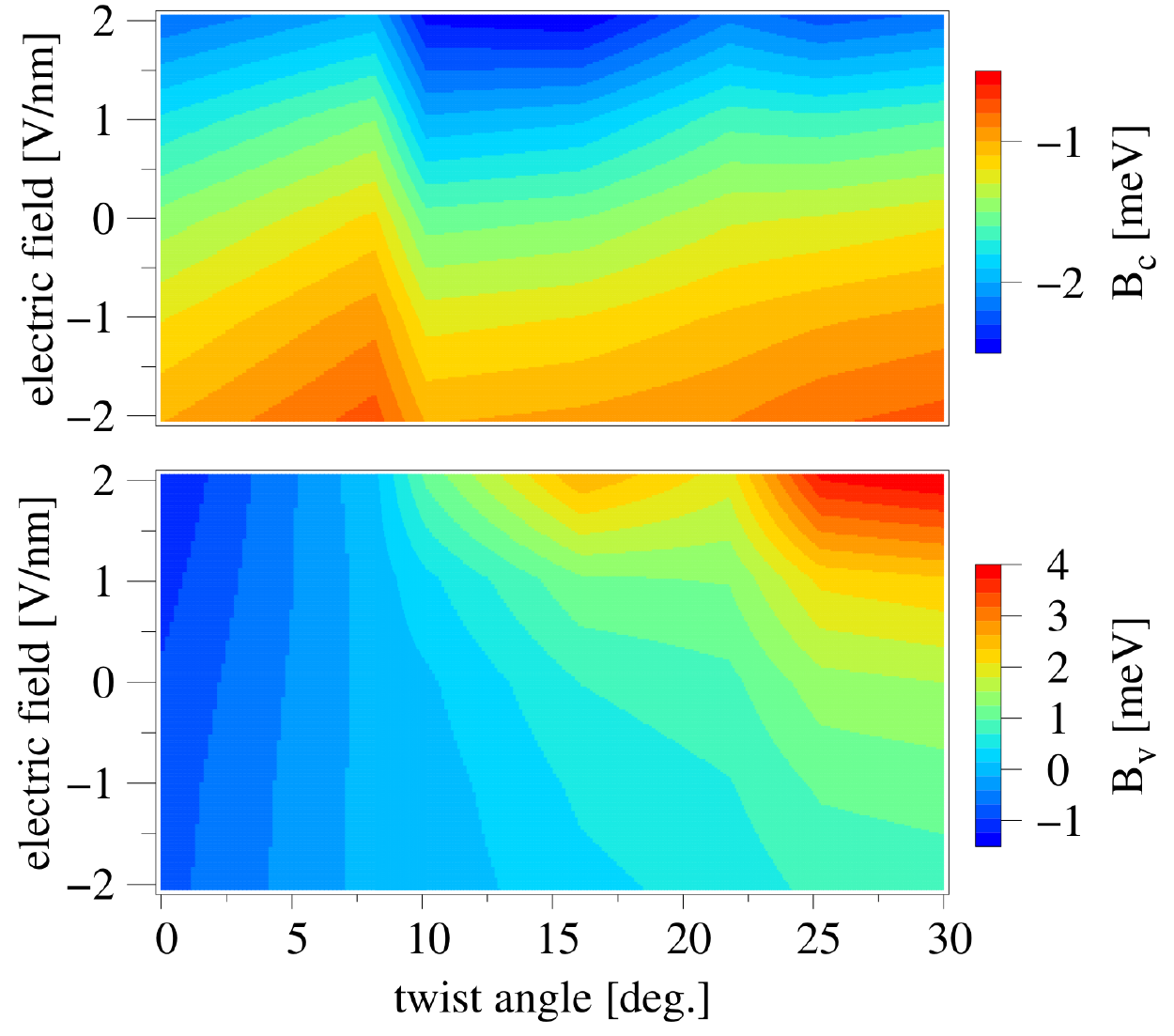}
     \caption{Calculated electric-field and twist-angle dependence of the proximity-induced exchange parameters $B_c$ (top) and $B_v$ (bottom) for the WSe$_2$/CrI$_3$ heterostructure, interpolated from Table~S2}
     \label{Fig:Efield}
    \end{figure}

In experimental geometries, gating is a useful tool to further control and tailor the proximity exchange coupling in TMDC/magnetic-semiconductor heterostructures. In the following, we consider the experimentally relevant  WSe$_2$/CrI$_3$ heterostructure~\cite{Zhong2017:SA,Seyler2018:NL,Zhong2020:NN} and perform a gate scan, in the range of $\pm 2$~V/nm, for all twist angles. The electric-field and twist-angle dependence of the proximity exchange parameters $B_c$ and $B_v$ is summarized in Fig.~\ref{Fig:Efield}, while the full fit results are summarized in Table~S2 \footnotemark[1]. For all angles and electric fields we consider, the CB proximity exchange parameter, $B_c$, stays negative and within the limits of about $-0.5$ to $-2.5$~meV. 
In contrast, the VB proximity exchange parameter, $B_v$, can be even stronger controlled by the electric field and the twist angle. As already mentioned, we find a crossover from positive to negative values at around $8^{\circ}$ twist angle for WSe$_2$. Furthermore, the applied field allows to tune $B_v$ in a wide range of values. Considering all angles, the tunability is giant and within the limits of about $-1.5$ to $4$~meV.

\subsection{Reversal of the valence band proximity exchange splitting}

The reversal of the TMDC VB edge splitting upon twisting is one of the most interesting findings, see Fig.~\ref{Fig:BC_BV_vs_angle}. 
A similar reversal of the proximity-induced exchange coupling has been recently found in
graphene/Cr$_2$Ge$_2$Te$_6$ bilayers when twisting from 0$^{\circ}$ to 30$^{\circ}$ \cite{Zollner2022:PRL}. The reversal was attributed to the different coupling of C orbitals to the spin up and spin down band manifolds of the magnetic substrate for the different twist angles. What is the mechanism for the investigated TMDC/CrI$_3$ bilayers?
For this purpose, we analyze the twisted WSe$_2$/CrI$_3$ heterostructures in more detail. In particular, we look at the three most relevant twist angles, namely 0$^{\circ}$, 8.2$^{\circ}$, and 30$^{\circ}$, corresponding to the cases when the WSe$_2$ VB edge proximity exchange splittings are negative, almost zero, and positive. A similar analysis should hold for MoSe$_2$, but the splitting vanishes at a different twist angle. 
In the following, we want to address the main points, while a more detailed discussion is given in the SM~\footnotemark[1].

    \begin{figure*}[htb]
     \includegraphics[width=.99\textwidth]{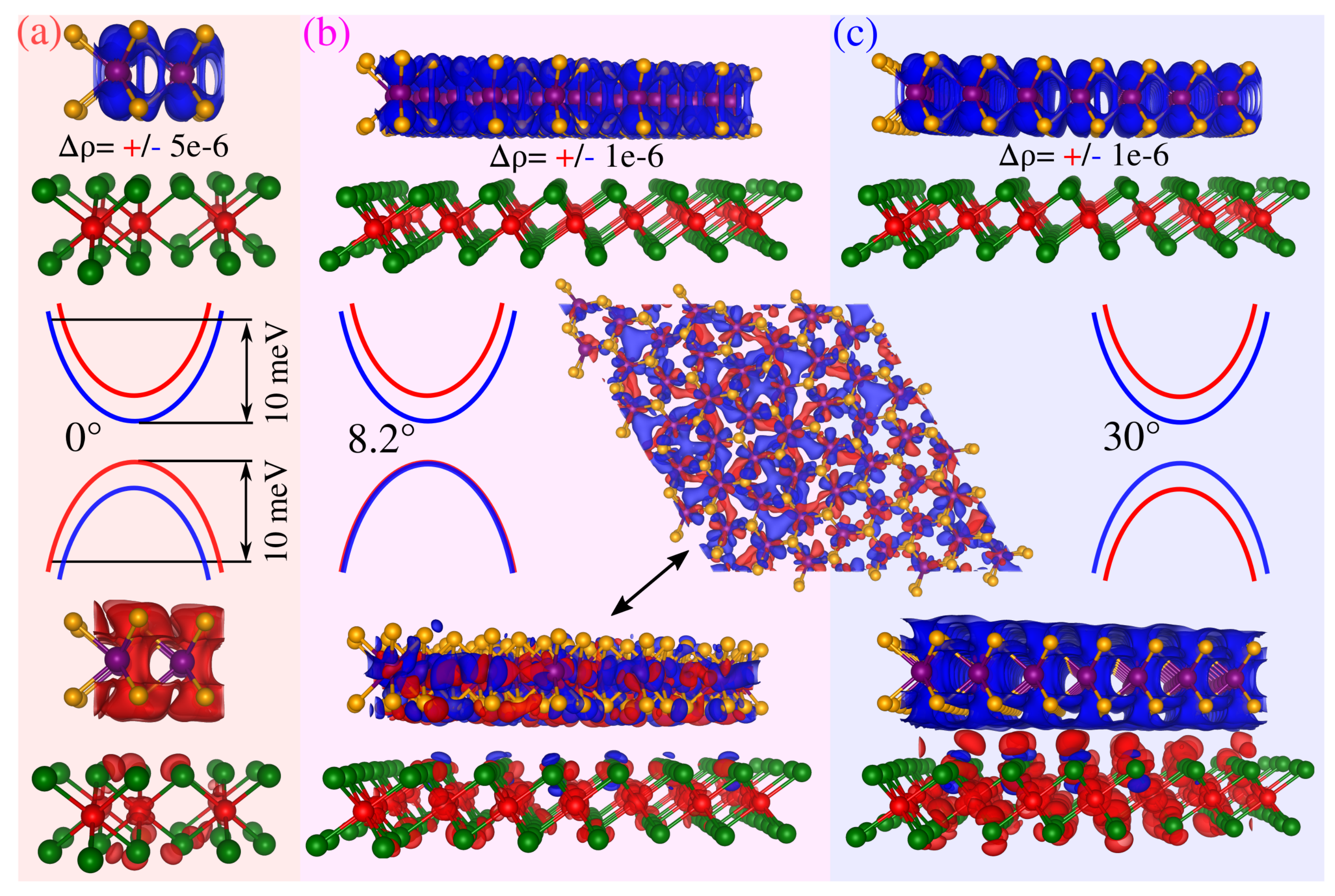}
     \caption{Calculated real space TMDC band edge spin polarizations, $\Delta\rho = \rho_{\uparrow}-\rho_{\downarrow}$, for the twisted WSe$_2$/CrI$_3$ heterostructures. The background color is used to group the subfigures for the three twist angles (a) 0$^{\circ}$, (b) 8.2$^{\circ}$, and (c) 30$^{\circ}$. 
     (a) Middle: Sketch of the proximitized TMDC band edges for 0$^{\circ}$, where we indicate the energy windows from which we calculate the spin densities $\rho_{\uparrow/\downarrow}$. Top/Bottom: Calculated band edge spin polarization, taking into account only CB/VB states. The color red (blue) corresponds to $\Delta\rho > 0$ ($\Delta\rho < 0$). The isosurfaces correspond to isovalues (units $\textrm{\AA}^{-3}$) as indicated. (b) The same as (a), but for 8.2$^{\circ}$. For the VB, we also show a top view, with removed CrI$_3$ layer, to show the non-uniform spin polarization on WSe$_2$. (c) The same as (a), but for 30$^{\circ}$.  }
     \label{Fig:spin_pol}
    \end{figure*}

To gain information about the proximity-induced exchange coupling, we look a the TMDC band edge spin polarizations in real space. Considering only TMDC CB edge states near the $K$ point, the calculated spin polarization is negative, independent of the twist angle and in agreement with the band edge dispersion, predominantly formed by W $d_{z^{2}}$ orbitals, and localized only on the WSe$_2$ layer (see Fig.~\ref{Fig:spin_pol}). In other words, the low energy TMDC CB edge states, which are spin-split due to proximity exchange, do not contain a spin polarization from the CrI$_3$ layer. Indeed, mainly in the spin down channel the high-energy TMDC and CrI$_3$ bands are coupled, as one can see in the projected band structures for the three twist angles (see Figs.~S3, S7, and S10). Considering second-order perturbation theory, the dominant coupling between spin down high-energy levels can repel the TMDC spin down band edge states to lower energies compared to the spin up ones, consistent with the observed splittings in Figs.~\ref{Fig:bands_MoSe2_CrI3}(b,e). Additionally considering the coupling to CrI$_3$ spin up in-gap states leads to the same conclusion. 

The explanation for the VB edge is a bit more involved. In Fig.~\ref{Fig:spin_pol}, we also show the calculated TMDC VB edge spin polarizations for 0$^{\circ}$, 8.2$^{\circ}$, and 30$^{\circ}$. 
In the case of 0$^{\circ}$, the VB edge splitting is negative [the dispersion is sketched in Fig.~\ref{Fig:spin_pol}(a)]. This leads to the positive spin polarization we can see in Fig.~\ref{Fig:spin_pol}(a), which is spread uniformly across the TMDC and predominantly formed by W $d_{xy}+d_{x^{2}-y^{2}}$ and Se $p$ orbitals. In addition, Cr $d_{zx}+d_{zy}$ and I $p_{x}+p_{y}$ orbitals contribute to the TMDC VB edge spin polarization.
For the 30$^{\circ}$ case, we find the opposite and a uniform negative band edge spin polarization on the TMDC, see Fig.~\ref{Fig:spin_pol}(c), again in agreement with the dispersion. 
For 8.2$^{\circ}$, the VB edge splitting almost vanishes and we find a highly non-uniform spin polarization, see Fig.~\ref{Fig:spin_pol}(b), with a multipole character around the W atoms. On average, the spin polarization is zero and therefore the band splitting is absent, but this does not rule out the complex spin structure we find.  
In fact, since the proximity exchange can sensitively depend on the stacking, as we have shown especially for the 0$^{\circ}$ supercell, one can expect local variations of the band edge spin polarizations also in experimentally realistic setups of the other twist angles.
Analyzing the TMDC VB states in more detail, there is apparently a delicate balance between the coupling to CrI$_3$ spin polarized in-gap and VB states. What is different is the backfolding of the TMDC $K$ point into the Brillouin zone of CrI$_3$ for different twist angles. For example, for $0^{\circ}$ the TMDC $K$ point folds back near the CrI$_3$ $K$ point, while at $30^{\circ}$ it folds back near the $\Gamma$ point, see Fig.~S5.
Since the interlayer coupling is different at different $k$ points, 
and also different CrI$_3$ band manifolds and orbitals play a role, 
this could indeed lead to the reversal of the TMDC VB edge splitting.

\section{Twist- and gate-tunable valley splitting}	
\label{sec:valley_splitting}

In order to provide insight into the optical signatures of the proximity exchange, we evaluate the valley splitting, i. e., the energy separation between the absorption peaks with opposite circular polarization ($\sigma^+$ and $\sigma^-$). Combining the proximity exchange parameters, $B_c$ and $B_v$, from Table~S2 \footnotemark[1], with the intrinsic SOC parameters for MoSe$_2$ and WSe$_2$ monolayers (see for example Ref.~\cite{Zollner2019:strain}), we can provide valuable insights regarding the twist-angle and electric-field dependence of the valley splitting. Besides the Hamiltonian terms given in Eq.~(\ref{Eq:Hamiltonian}), the SOC term reads
\begin{equation}
H_{\text{soc}} = \tau s_{z}\otimes\left(\lambda_{\text{c}}\sigma_{+}+\lambda_{\text{v}}\sigma_{-}\right) \, ,
\label{eq:Hsoc}
\end{equation}
with parameters $\lambda_{\text{v}} = 94.56\; \textrm{meV}$ and $\lambda_{\text{c}} = -9.647\; \textrm{meV}$ for MoSe$_2$, and $\lambda_{\text{v}} = 241.79 \; \textrm{meV}$ and $\lambda_{\text{c}} = 13.9\; \textrm{meV}$ for WSe$_2$, taken from Ref.~\cite{Zollner2019a:PRB}.

To incorporate excitonic effects and calculate the absorption spectra, we apply the robust formalism of the effective Bethe-Salpether equation with the electron-hole Coulomb interaction mediated by the Rytova-Keldysh potential, similar to Ref.~\cite{Zollner2019a:PRB,Zollner2020:PRB}. From the calculated absorption, we can extract the valley splitting, i. e., the energy difference between the first A exciton peaks at $K$ and $K'$ (with opposite circular polarization). For the excitonic calculations, the BSE is solved numerically using 101$\times$101 k-points in a square region with sides ranging from $-0.5$ to 0.5 $\textrm{\AA}^{-1}$, thus leading to a k-point spacing of $\Delta k = 10^{-2} \; \textrm{\AA}^{-1}$ along each direction. Additionally, the Coulomb potential is averaged in a submesh of 101$\times$101 points covering an area of $\Delta k^2$ around each k-point. The screening length used in the calculations is $r_0 = 51~(45)~\textrm{\AA}$ for MoSe$_2$ (WSe$_2$), from Ref.~\cite{Berkelbach2013:PRB}, and the dielectric constant for CrI$_3$ is taken as $\epsilon = 1.8$, from Ref.~\cite{Huang2017:Nat}.

In the single-particle limit, i. e., where no excitonic effects are taken into account, the valley splitting can be obtained analytically by solving the Hamiltonians (\ref{Eq:Hamiltonian}) and (\ref{eq:Hsoc}) and is given by

\begin{align}
\text{VS} & =\left(E_{\text{c}\uparrow,\text{K}}-E_{v\uparrow,\text{K}}\right)-\left(E_{\text{c}\downarrow,\text{-K}}-E_{\text{v}\downarrow,\text{-K}}\right)\nonumber \\
 & =-2\left(B_{\text{c}}-B_{\text{v}}\right) \, .
 \label{eq:VSsp}
\end{align}

\begin{figure}[htb]
\includegraphics[width=1.0\columnwidth]{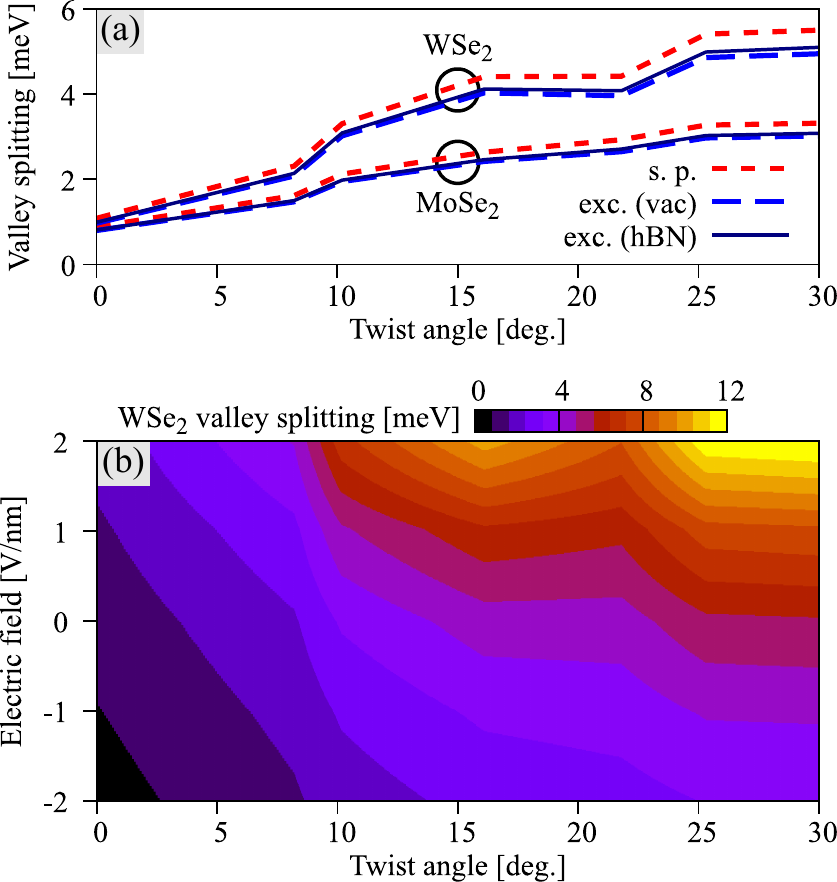}
\caption{(a) Valley splitting for the MoSe$_2$/CrI$_3$ and WSe$_2$/CrI$_3$ bilayers calculated from the single-particle (abbreviated by s.p., short dashed lines), and from the excitonic absorption (abbreviated as exc.) considering the region above the TMDC to be vacuum (long dashed lines) or hexagonal boron nitride (solid lines). The single-particle values follow nicely the excitonic calculations. (b) Valley splitting for the WSe$_2$/CrI$_3$ bilayer as function of the twist angle (x-axis) and applied electric field (y-axis) using the single-particle values, calculated via Eq.(\ref{eq:VSsp})}
\label{Fig:VS}
\end{figure}

In Fig.~\ref{Fig:VS}(a) we reveal the behavior of the valley splitting as a function of the twist angle for MoSe$_2$/CrI$_3$ and WSe$_2$/CrI$_3$. For both cases, the valley splitting increases as the twist angle gets larger, almost in a linear fashion. 
The role of excitonic effects due to different environments on top of the TMDC is also explored: air/vacuum with $\varepsilon=1$ (long dashed lines) and hexagonal boron nitride with $\varepsilon=4.5$~\cite{Geick1966PR,Stier2018:PRL,Goryca2019NatComm} (solid lines). The single-particle values (short dashed lines), obtained from Eq.~(\ref{eq:VSsp}), are also shown and follow closely the trends of the excitonic valley splitting extracted from the absorption spectra (shown in the Supplemental Material \footnotemark[1]), suggesting that it is possible to grasp valuable insight into the valley splitting just by knowing the exchange parameters $B_c$ and $B_v$ extracted from DFT calculations. Our results reveal that the twist-angle is a crucial parameter for the assembly of TMDC/ferromagnetic-semiconductor vdW heterostructures and, particularly, it has a sizable impact on the magnitude of the valley splitting.

To highlight the extraordinary tunability of the valley splitting by the combined effect of the twist angle and gating, we show in Fig.~\ref{Fig:VS}(b) a complete color map for the WSe$_2$/CrI$_3$ system, using the single-particle valley splitting expression of Eq.~(\ref{eq:VSsp}). On the bottom left corner of the figure, the valley splitting nearly vanishes for 0$^\circ$ twist angle and $\sim -2$~V/nm, whereas on the top right corner, the valley splitting reaches very large values of approximately 12~meV for 30$^\circ$ and $\sim 2$~V/nm. For a fixed twist angle, the dependence of the valley splitting is almost linear, in agreement with our previous calculations~\cite{Zollner2019a:PRB} and experimentally demonstrated in MoSe$_2$/CrBr$_3$ samples~\cite{Ciorciaro2020:arxiv}.

\section{Summary}	
\label{sec:summary}

We have shown that, by employing first-principles calculations on large supercells, one can tailor the proximity exchange coupling in TMDC/CrI$_3$ bilayers by twisting and gating. The previously observed
reversal of the proximity exchange is demonstrated to result from a systematic, essentially linear dependence of the exchange in the VB on the twist angle. We do not find any significant variation in the CB. 

Analyzing the real space spin-resolved electronic density for the band edges, we find
that the CB electrons exhibit a uniform spatial spin polarization for all the investigated twist angles. In contrast, the hole spin polarization undergoes remarkable transformations as going from 0 to 30 degrees. While at the two limit angles the spin polarizations are highly uniform, albeit opposite (corresponding to the opposite spin splittings), at the twist 
angles where the spin splitting reverses sign the spin polarization is highly non-uniform, 
alternating between spin up and spin down orbitals, arising from a subtle balance between 
orbital- and spin-resolved hybridization of the TMDC and CrI$_3$ orbitals. 
By comparing different stacking configurations, we have also demonstrated that the induced magnetic moments, proximity exchange, and the spin polarizations can vary locally in space within experimentally realistic setups. 

Further substantial tunability of the band edge splittings is provided by a transverse electric field of a few V/nm. 
A low-energy model Hamiltonian, with fitted parameters that nicely reproduce the relevant DFT-calculated band edges of the proximitized TMDCs, has been employed to reveal 
the experimentally verifiable signatures of the CrI$_3$ exchange field on the valley splitting of the first TMDC exciton peak. Particularly, we predict that the valley splitting increases linearly with respect to the twist angle.
Furthermore, in WSe$_2$/CrI$_3$ heterostructures the valley splitting shows an unprecedented high tunability upon twisting and gating, ranging from 0 to 12~meV, equivalent to external magnetic fields of up to 60 Tesla. 
Our results also stress the importance of documenting the twist angle when employing magnetic vdW heterostructures in experiments.

\acknowledgments
This work was funded by the Deutsche Forschungsgemeinschaft (DFG, German Research Foundation) SFB 1277 (Project No. 314695032), SPP 2244 (Project No. 443416183), and the European Union Horizon 2020 Research and Innovation Program under contract number 881603 (Graphene Flagship).
    
\footnotetext[1]{See Supplemental Material, including Refs. \cite{Zollner2022:PRL,Zollner2019:strain,Zollner2019a:PRB,David2019:arxiv,Li2019:PRB,McGuire2015:CM, Schutte1987:JSSC, James1963:AC,Wu2019:PCCP,FariaJunior2022}, where we show all twisted heterostructure geometries and discuss about the origin of the reversal of the TMDC VB spin splitting. 
In addition, we list the model parameters for the electric field study, briefly address different stackings, the influence of strain, the twist angles between $30^{\circ}$ to $60^{\circ}$, and show the calculated absorption spectra. }

\bibliography{references}

\cleardoublepage


\section*{Supplementary material}

# Strong manipulation of the valley splitting upon twisting and gating in MoSe$_2$/CrI$_3$ and WSe$_2$/CrI$_3$ van der Waals heterostructures

Klaus Zollner,[1] Paulo E. Faria Junior,[1] and Jaroslav Fabian[1]

[1]*Institute for Theoretical Physics, University of Regensburg, 93040 Regensburg, Germany*

## I. ATOMIC CONFIGURATIONS

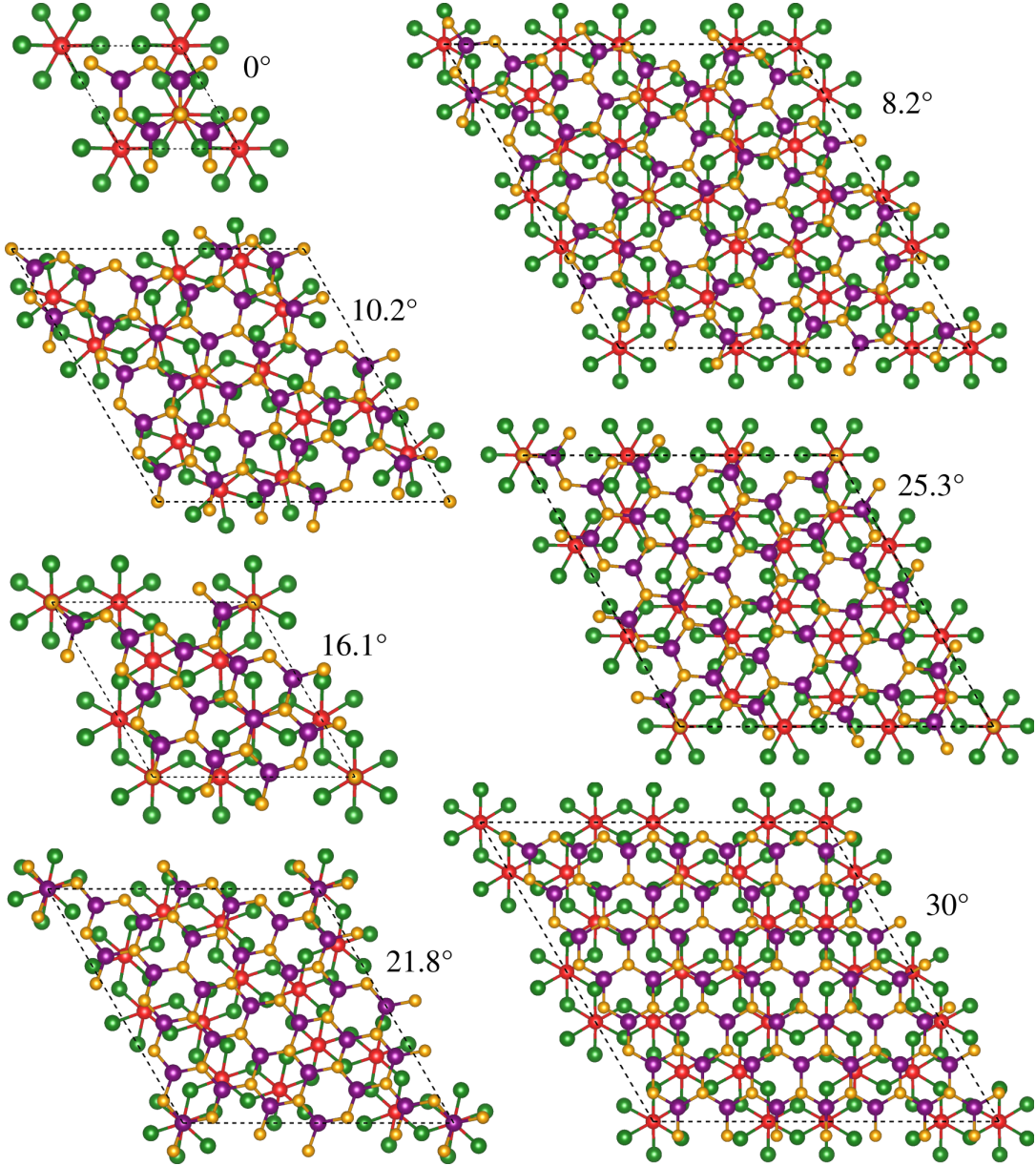

FIG. S1. Top views of the twisted TMDC/CrI$_3$ geometries. The dashed lines are the edges of the unit cell. Cr = red, I = green, Mo/W = purple, and Se = orange.

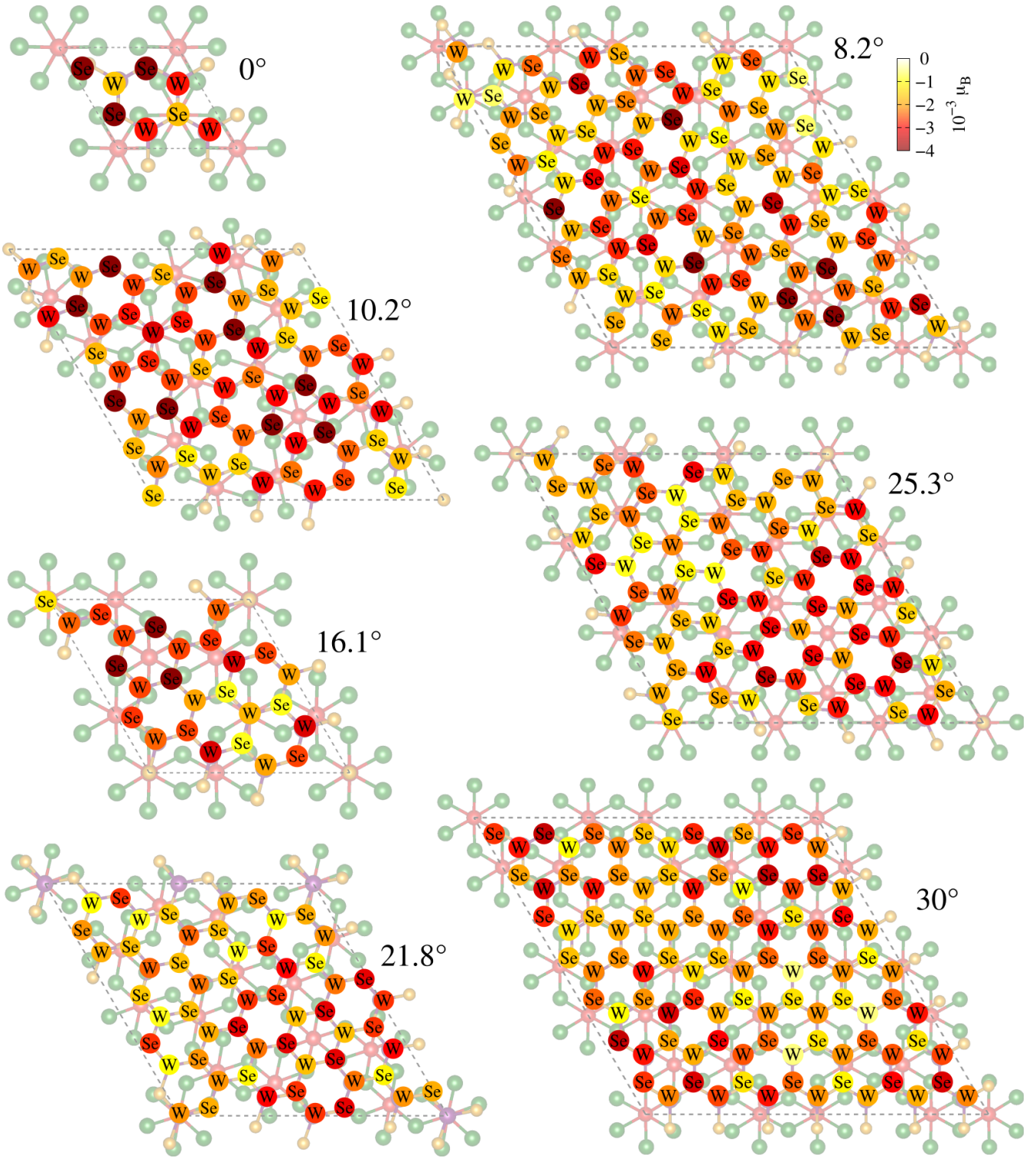

FIG. S2. Same as Fig. S1, where we overlay each structure with colored spheres that represent the proximity-induced local magnetic moments of the interfacial Se and W atoms.

## II. ORIGIN OF PROXIMITY EXCHANGE

The reversal of the TMDC valence band edge splitting upon twisting is the most interesting finding of our investigations. A similar reversal of the proximity-induced exchange coupling has been recently found in graphene/$Cr_2Ge_2Te_6$ bilayers when twisting from 0° to 30° [1]. The reversal was attributed to the different coupling of C orbitals to the spin up and spin down band manifolds of the magnetic substrate for the different twist angles. In order to understand this reversal in the case of our TMDC/$CrI_3$ bilayers, one first has to know that the TMDC valence (conduction) band edge at $K$ is predominantly formed by transition-metal $d_{xy+x^2-y^2}$ ($d_{z^2}$) orbitals with some amount of chalcogen $p$ orbitals [2]. Since the coupling of the different orbitals to the substrate may be different, this can lead to different splittings of the band edges. One way to find out about the couplings is within the frame of the projected band structure.

In Fig. S3, we show the DFT-calculated projected band structure of the $WSe_2$/$CrI_3$ heterostructures for a twist angle of 0° decomposed into spin up and spin down band manifolds. We find that in the spin up manifold, only the TMDC valence band states are heavily coupled to $CrI_3$ bands, as indicated by the yellow and green colors. The spin up in-gap states from $CrI_3$ do not seem to contribute in the coupling. Moreover, there are no spin up states of $CrI_3$ within the TMDC conduction band states. In the spin down manifold, both TMDC valence and conduction band states are coupled to $CrI_3$. However, the TMDC band edge states at $K$ do not show a major contribution of $CrI_3$ and the coupling happens mainly between the high-energy levels. Since the conduction band states are only coupled in the spin down manifold and second-order perturbation theory predicts level repulsion, the TMDC spin down conduction band edge states are shifted to lower energies compared to spin up ones. This is indeed consistent with the spin splitting of the low energy band structure for 0°. Actually, this is valid for all twist angles. Looking at the projected band structures for 8.2°, see Fig. S7, and 30°, see Fig. S10, $CrI_3$ conduction band states are only present in the spin down manifold, pushing the TMDC spin down conduction band edge states at $K$ to lower energies. This is a consequence of the monolayer $CrI_3$ dispersion, see Fig. S5, having only spin down states at higher energies and the specific band alignment within the TMDC/$CrI_3$ heterostructure. Moreover, for 0° the TMDC band edge states at the $K/K'$ points are backfolded near the $K$ points of $CrI_3$, see the inset in Fig. S3. Analyzing the monolayer $CrI_3$ band structure, see Fig. S5, and looking at the $k$ point to which the TMDC band edge states couple to for 0°, spin down conduction bands are almost exclusively formed by Cr $d$ orbitals. To get an even clearer picture, we have also calculated the atom and spin resolved density of states (DOS), see Fig. S4 for 0° twist angle. We find that also the $CrI_3$ in-gap states contain some small contribution of $WSe_2$, which is not obvious from the projected band structure. Conduction band states of $CrI_3$ and $WSe_2$ are coupled in the spin down manifold only, as we have already seen in Fig. S3. Taking into account also the coupling to the $CrI_3$ spin up in-gap states, still the TMDC spin up conduction band edge states would be shifted to higher energies compared to spin down ones, which is again consistent with the observed band edge splittings. The above analysis nicely explains the rather constant splitting of the TMDC conduction band edge upon twisting.

What about the valence band edge splitting? From the monolayer $CrI_3$ dispersion, see Fig. S5, we find that the valence bands are formed mainly by I $p$ orbitals, while in-gap states are formed also by Cr $d$ orbitals. From the projected band structure for 0°, see Fig. S3, we see that both spin species contribute in the coupling of valence bands. However, we have also found from the DOS, see Fig. S4, that spin up in-gap states contribute in the coupling. This picture holds for the other twist angles as well, see Figs. S7, S10, S8, and S11. If one would consider only a coupling to the spin up in-gap states, TMDC spin up valence band edge states would be always shifted to lower energies compared to spin down ones, according to level repulsion from second-order perturbation theory. However, this is not what we observe for the different twist angles. Apparently, for the TMDC valence band edge splitting, there is a delicate balance between the coupling to $CrI_3$ spin polarized in-gap and valence band states. What is different is the backfolding of the TMDC $K$ point into the Brillouin zone of $CrI_3$ for different twist angles (see the insets in the projected band structures). For example, for 0° the TMDC $K$ point folds back near the $CrI_3$ $K$ point, while at 30° it folds back near the $\Gamma$ point, see also Fig. S5. Since the interlayer coupling is different at different $k$ points, and also different $CrI_3$ band manifolds and orbitals play a role, this could indeed lead to the reversal of the TMDC valence band edge splitting. Here, we can only give indications and, unfortunately, no definite answer on this.

We have also calculated the spin polarizations in real space, $\Delta\rho = \rho_\uparrow - \rho_\downarrow$, where the spin densities $\rho_{\uparrow/\downarrow}(\boldsymbol{r}) = \sum_{n,k} |\phi_{n,\uparrow/\downarrow}^{\boldsymbol{k}}(\boldsymbol{r})|^2$ are sums over eigenstates with the corresponding spin. To find out about the proximity exchange couplings, we take into account the TMDC band edge states near the $K$ point within an energy window of 10 meV, as indicated by the sketch in Fig. S6, for the calculation of spin densities (conduction and valence bands individually). For the 0° $WSe_2$/$CrI_3$ heterostructure (Fig. S6), we find that the conduction band edge states have a $d_{z^2}$ orbital character and the spin down density outweighs the spin up one, as expected from the spin-split band edge dispersion. However, there is no coupling to any magnetic substrate states, since the band edge spin densities contain no $CrI_3$ contributions. In contrast, for the valence band edge, the spin up density outweighs the spin down one. Moreover, we find a significant coupling to the $CrI_3$ spin up density. In particular, Cr $d_{zx+zy}$ and I $p_{x+y}$ orbitals contribute to the

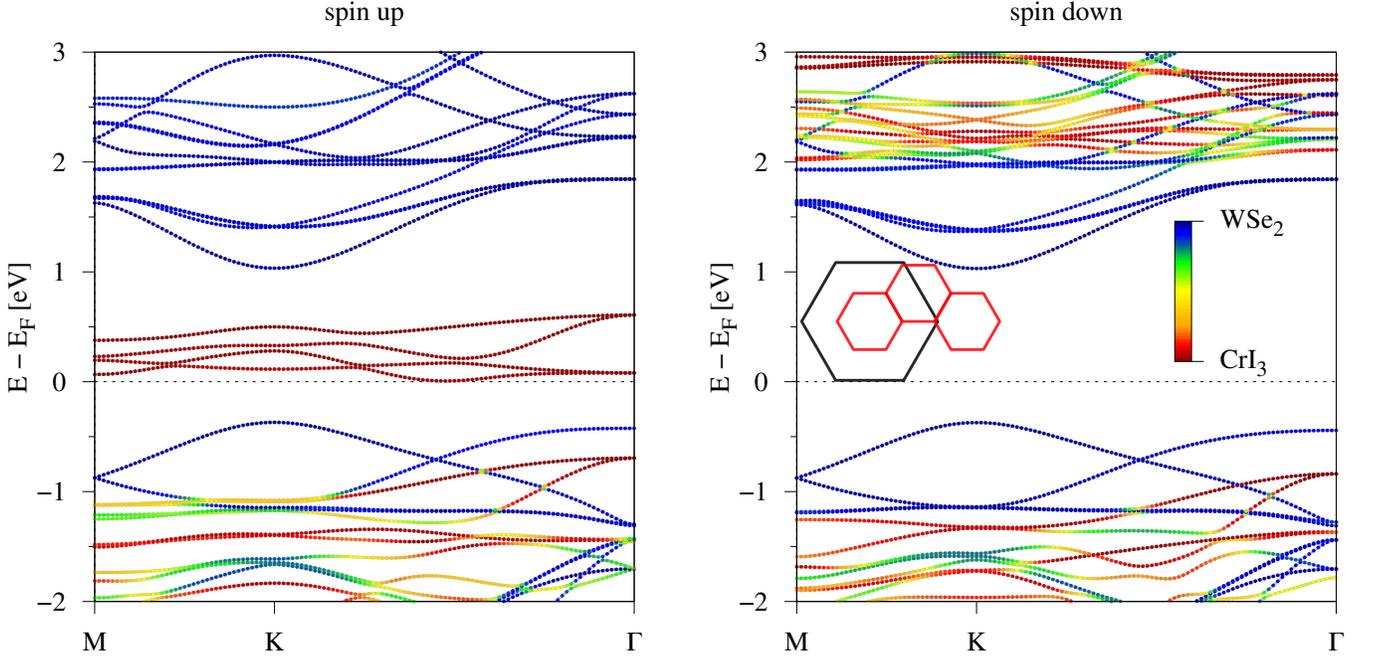

FIG. S3. DFT-calculated projected band structure of the WSe$_2$/CrI$_3$ heterostructures for a twist angle of 0°. Left: Spin up bands. Right: Spin down bands. The color code shows the contribution of the individual monolayers to the bands, i. e., the bands appear dark-reddish (dark-blueish) when only CrI$_3$ (WSe$_2$) orbitals contribute. The inset in the spin down manifold shows the backfolding of the TMDC $K/K'$ points. The black (red) hexagon represents the TMDC (CrI$_3$) Brillouin zone. The sizes of the Brillouin zones are scaled according to the experimental lattice constants.

TMDC valence band edge, which is predominantly formed by W $d_{xy+x^2-y^2}$ orbitals.

The band edge spin polarizations for the 8.2° structure are shown in Fig. S9. Remember that for this angle, the WSe$_2$ valence band edge splitting vanishes. Again, the TMDC conduction band edge does not contain any CrI$_3$ states. The spin polarization is negative, has a $d_{z^2}$ orbital character and is spread uniform across the TMDC, just as for 0°. Looking at the valence band edge spin polarization, we find it to be highly non-uniform with a multipole character around the W atoms. On average, the spin polarization is zero and therefore the band splitting is absent, but this does not rule out that a complex spin structure can arise. Moreover, again Cr $d_{zx+zy}$ and I $p_{x+y}$ orbitals contribute to the TMDC valence band edge.

Also for the 30° structure, we have calculated the band edge spin polarizations, see Fig. S12. Similar to 0°, the conduction band edge contains no CrI$_3$ states, and W $d_{z^2}$ orbitals are dominant. Within the valence band edge, there is again a sizable spin up density from the CrI$_3$. Comparing the TMDC valence band edge spin polarizations in Fig. S6 for 0° and Fig. S12 for 30°, we find them to be opposite, in agreement with the band structures.

TABLE S1. Calculated averaged atomic magnetic moments in units of $\mu_B$ of the twisted MoSe$_2$/CrI$_3$ (WSe$_2$/CrI$_3$) heterostructures. For Se, we consider only the atoms at the interface.

| $\vartheta$ [°] | Mo (W) | Se | Cr | I |
|---|---|---|---|---|
| 0.0000 | -0.0025 (-0.0026) | -0.0032 (-0.0034) | 3.3517 (3.3511) | -0.1496 (-0.1491) |
| 8.2132 | -0.0019 (-0.0020) | -0.0022 (-0.0023) | 3.3520 (3.3529) | -0.1500 (-0.1500) |
| 10.1583 | -0.0025 (-0.0026) | -0.0026 (-0.0027) | 3.3968 (3.3993) | -0.1591 (-0.1594) |
| 16.1021 | -0.0022 (-0.0025) | -0.0024 (-0.0026) | 3.3772 (3.3772) | -0.1551 (-0.1547) |
| 21.7868 | -0.0018 (-0.0020) | -0.0022 (-0.0022) | 3.3440 (3.3448) | -0.1484 (-0.1484) |
| 25.2850 | -0.0019 (-0.0022) | -0.0021 (-0.0023) | 3.3544 (3.3558) | -0.1506 (-0.1506) |
| 30.0000 | -0.0019 (-0.0021) | -0.0021 (-0.0022) | 3.3529 (3.3547) | -0.1502 (-0.1503) |

In summary, from our DFT analysis we find no clear argument why the reversal of the valence band spin splitting in the TMDC occurs upon twisting. The main reason should be the twist-angle dependent backfolding of the TMDC band edges at $K/K'$ to different $k$ points within the CrI$_3$ Brillouin zone, where atomic orbitals contribute differently to the interlayer coupling. In addition, one probably has to take into account that the magnetic moments within the

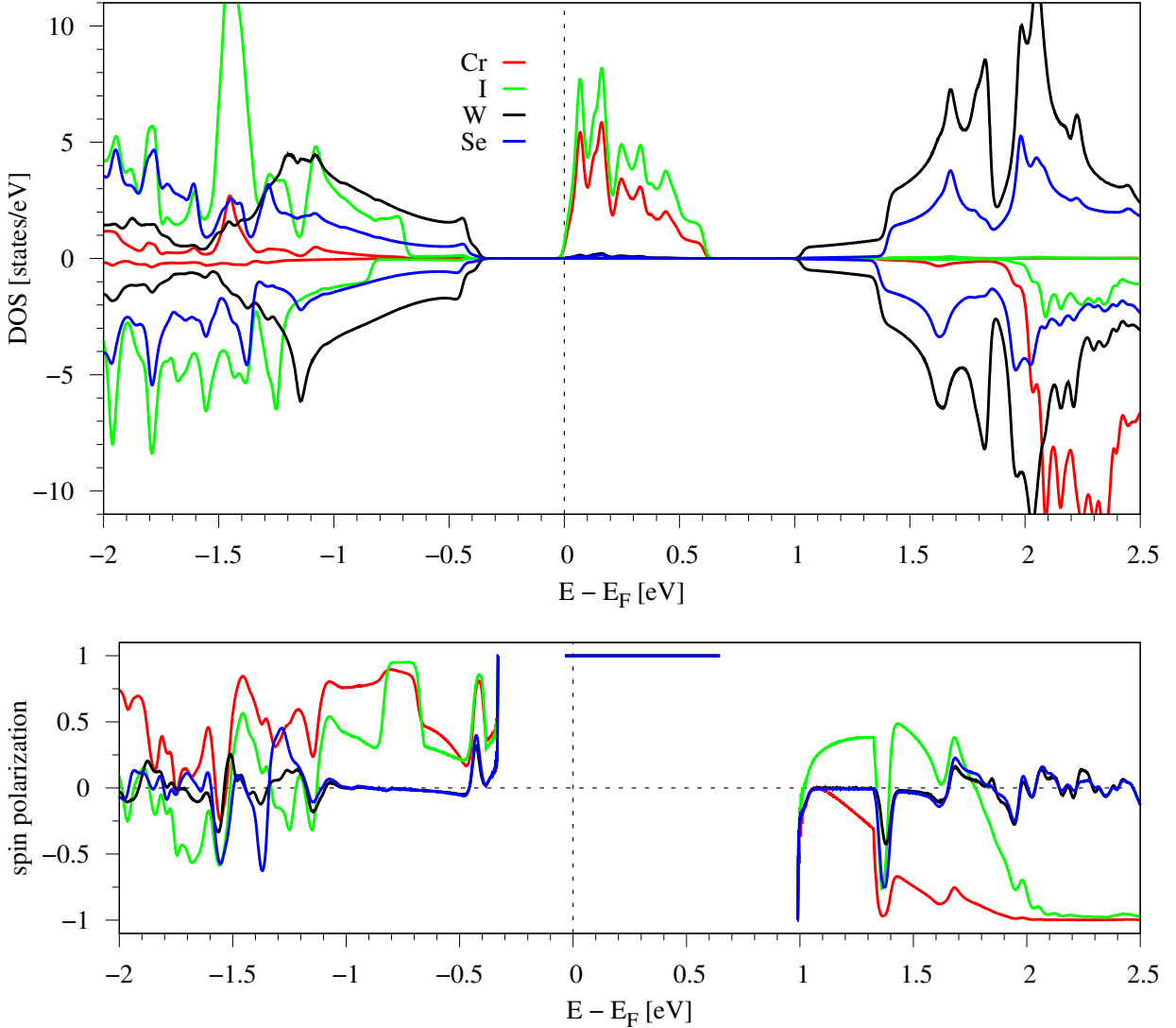

FIG. S4. Top: DFT-calculated spin and atom resolved density of states (DOS) of the WSe$_2$/CrI$_3$ heterostructures for a twist angle of 0°. Positive (negative) DOS corresponds to spin up (spin down), while the different colors correspond to different atoms. Bottom: The corresponding spin polarizations calculated as $(N_\uparrow - N_\downarrow)/(N_\uparrow + N_\downarrow)$, where $N_\uparrow$ ($N_\downarrow$) is the spin up (spin down) DOS of the different atoms.

CrI$_3$ substrate are opposite for Cr ($\sim 3.35\ \mu_B$) and I ($\sim -0.15\ \mu_B$) atoms, and the induced magnetic moments in the TMDC are parallel to the ones of I atoms, see Table S1, nearly independent of the twist angle. To clarify the origin of this reversal, one may look at the coupled bilayer system within the tight-binding framework, similar as in Refs. [3, 4] for graphene/TMDC systems.

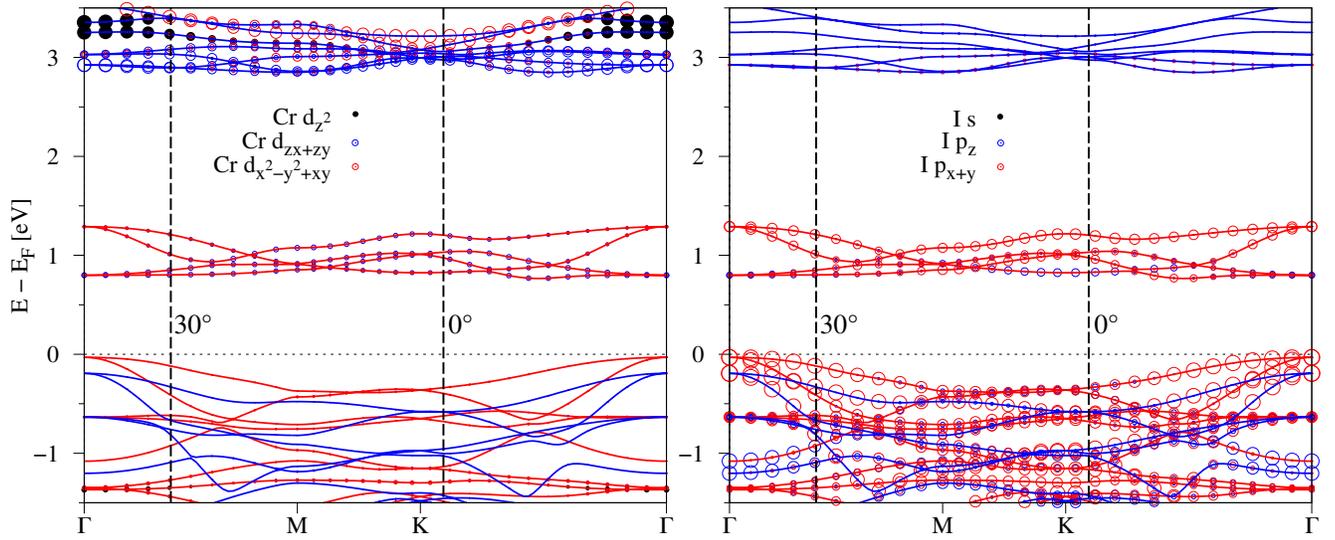

FIG. S5. DFT-calculated band structure of monolayer CrI$_3$. Left: projected onto Cr $d$ orbitals. Right: projected onto I $s$ and $p$ orbitals. The size of the symbols represent the contribution of the orbitals to a specific band and at a given $k$ point. The color of the solid lines represent spin up (red) and spin down (blue). The vertical dashed lines indicate the backfolding of the TMDC $K$ point, into the CrI$_3$ Brillouin zone, according to the insets in Fig. S3 and Fig. S10.

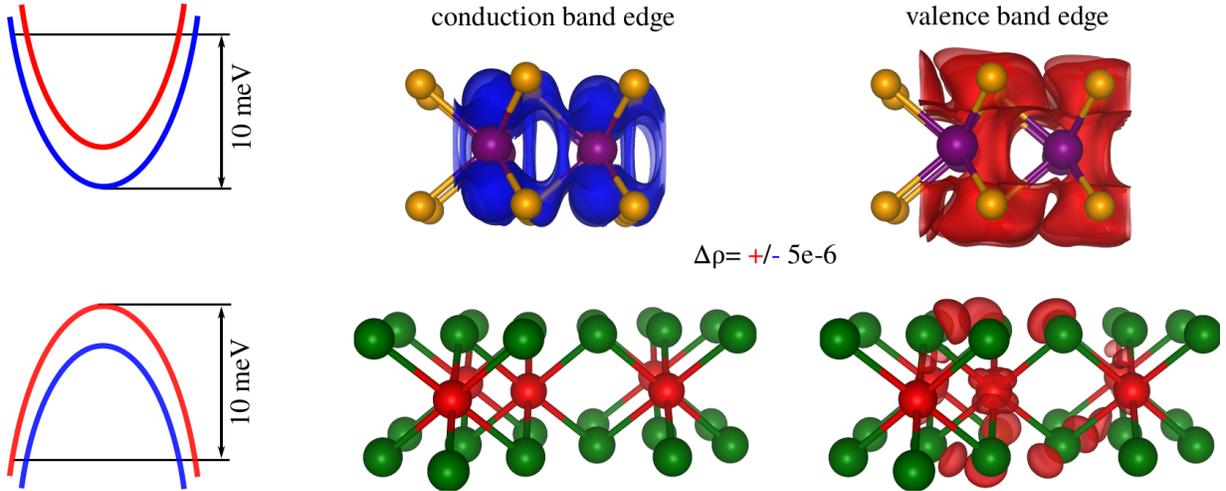

FIG. S6. Calculated spin polarization, $\Delta\rho = \rho_\uparrow - \rho_\downarrow$, for the WSe$_2$/CrI$_3$ heterostructure with a twist angle of 0°. The color red (blue) corresponds to $\Delta\rho > 0$ ($\Delta\rho < 0$). The isosurfaces correspond to isovalues (units Å$^{-3}$) as indicated. Middle (right) figure takes into account WSe$_2$ conduction (valence) band edge states near the $K$ point within an energy window of 10 meV, as indicated by the left sketch of the spin-split TMDC band edges.

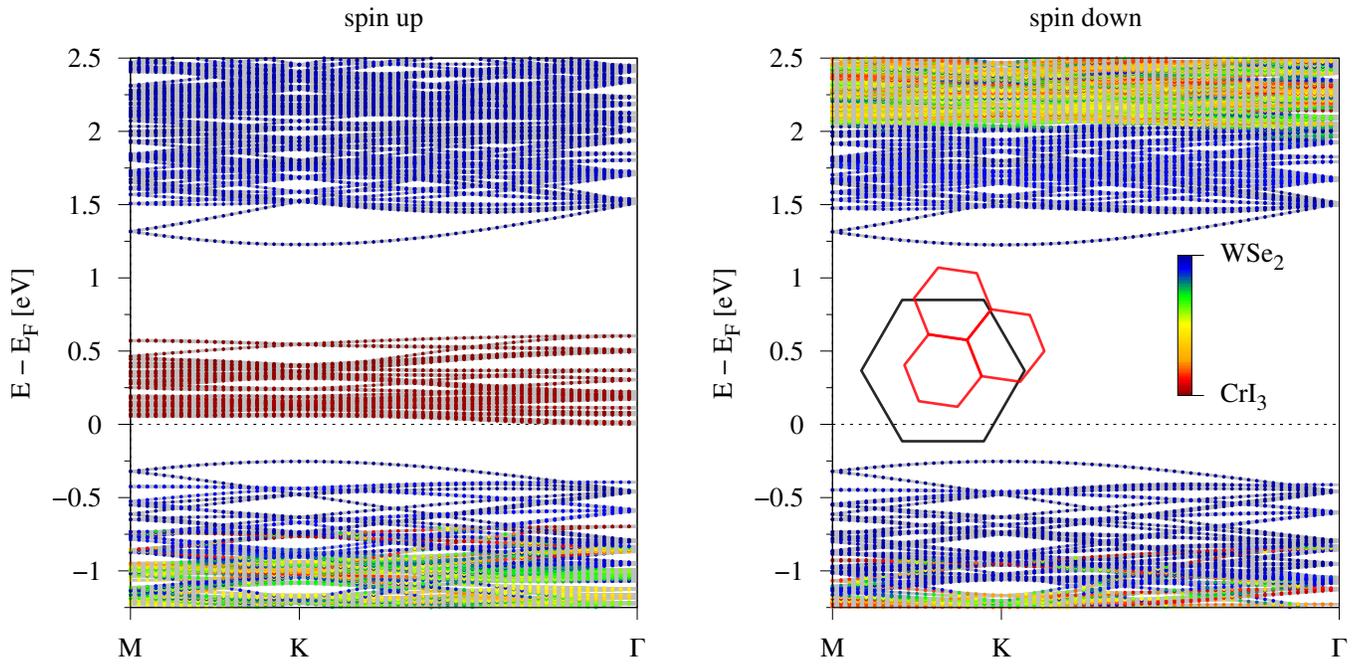

FIG. S7. Same as Fig. S3, but for a twist angle of 8.2°

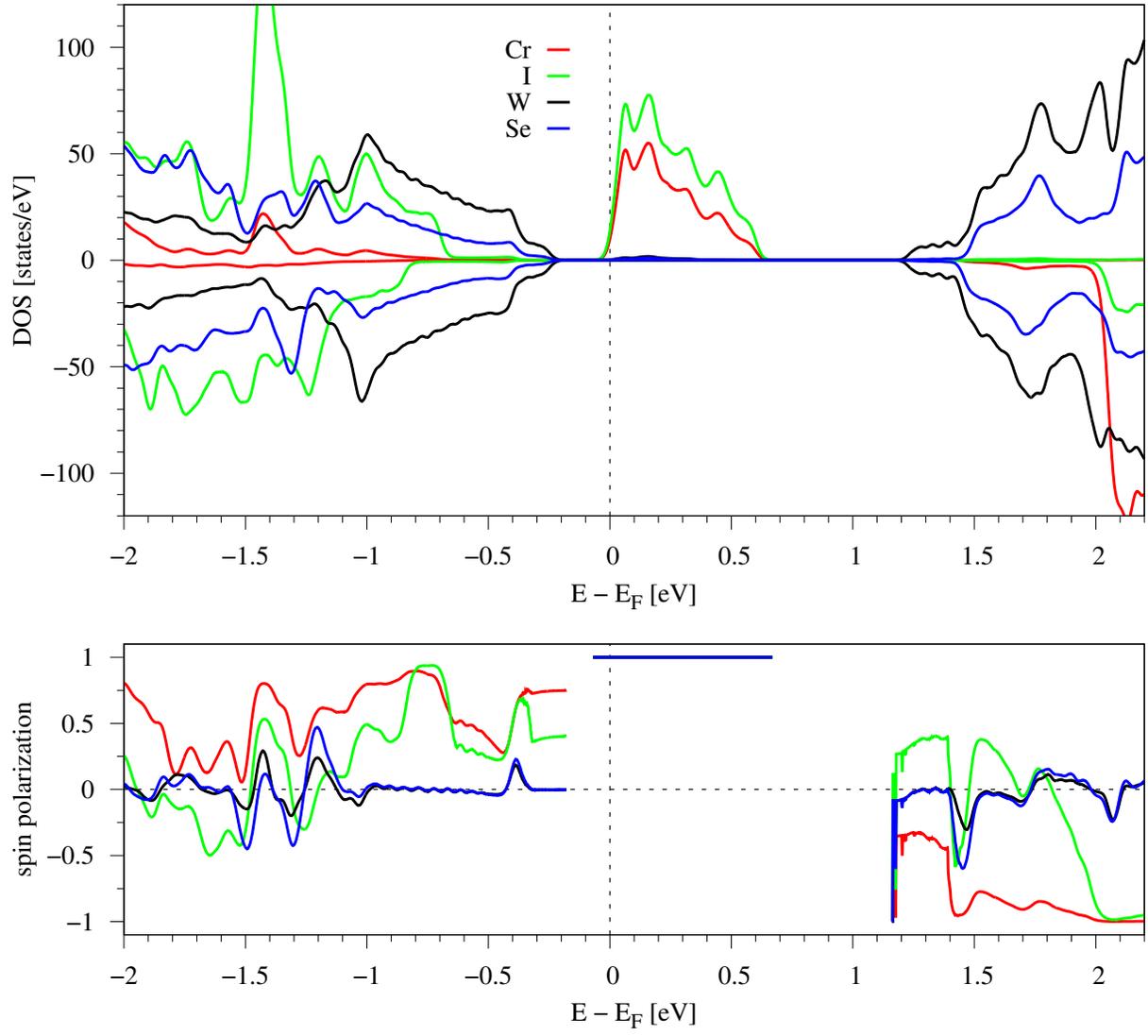

FIG. S8. Same as Fig. S4, but for a twist angle of 8.2°.

conduction band edge     $\Delta\rho = $ +/- 1e-6     valence band edge

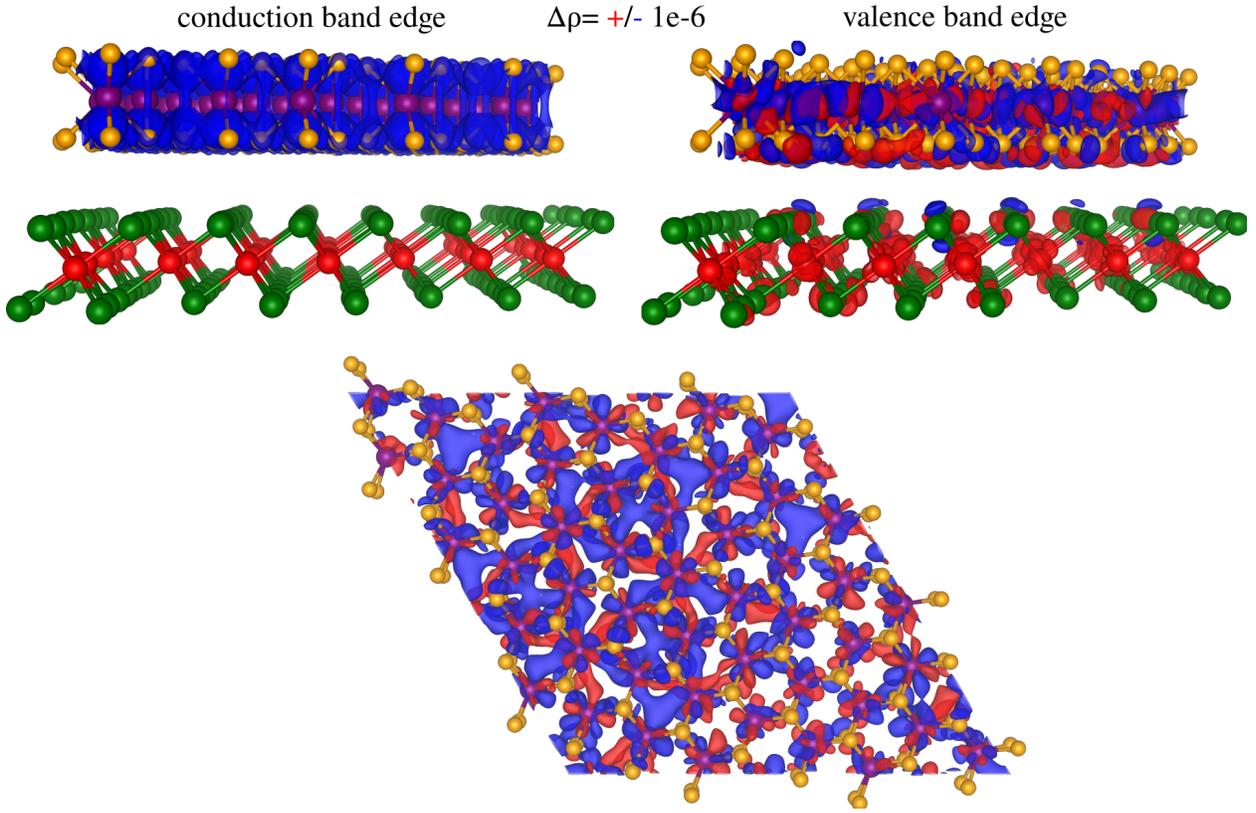

FIG. S9. Calculated spin polarization, $\Delta\rho = \rho_\uparrow - \rho_\downarrow$, for the WSe$_2$/CrI$_3$ heterostructure with a twist angle of 8.2°. The color red (blue) corresponds to $\Delta\rho > 0$ ($\Delta\rho < 0$). The isosurfaces correspond to isovalues (units Å$^{-3}$) as indicated. Top left (right) figure takes into account WSe$_2$ conduction (valence) band edge states near the $K$ point within an energy window of 10 meV, similar as in Fig. S6. Bottom figure is a top view for the valence band edge without the CrI$_3$ layer, to show the highly non-uniform spin polarization on WSe$_2$.

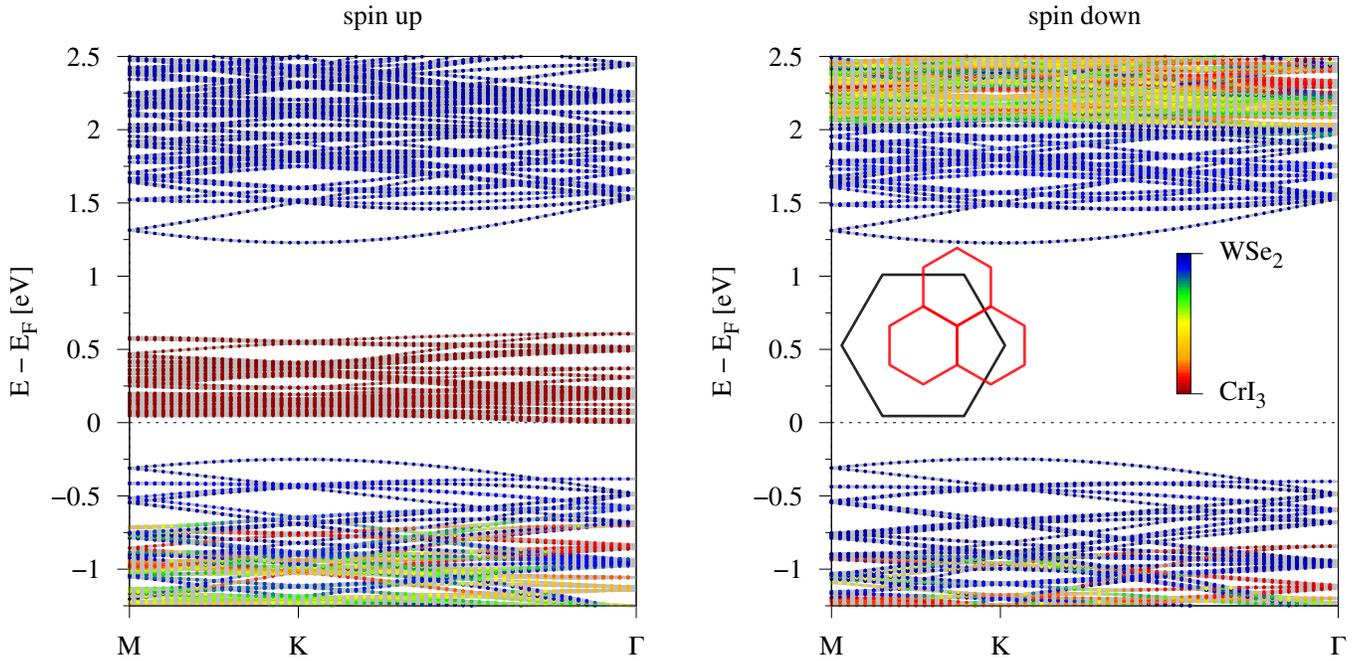

FIG. S10. Same as Fig. S3, but for a twist angle of 30°.

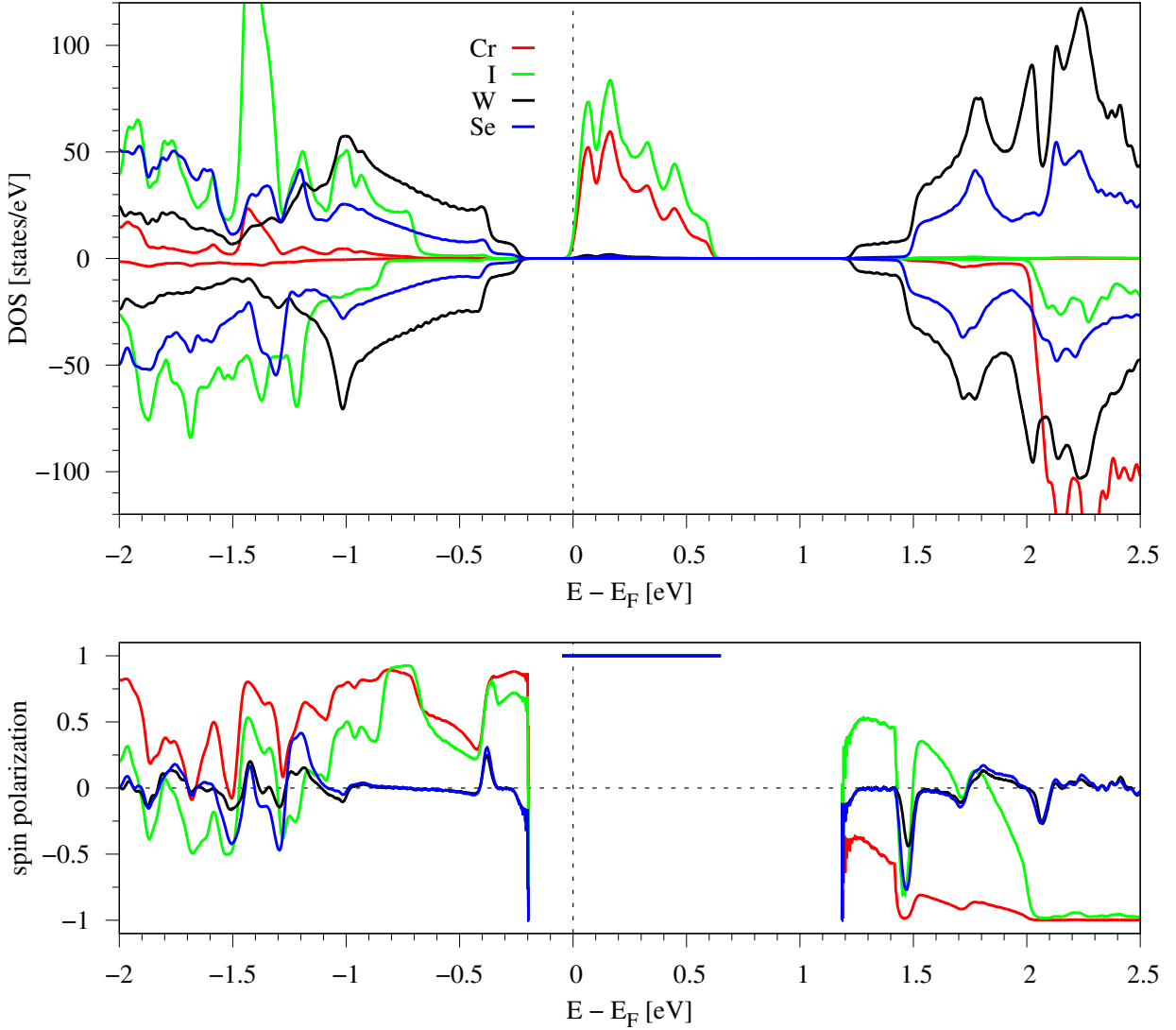

FIG. S11. Same as Fig. S4, but for a twist angle of 30°.

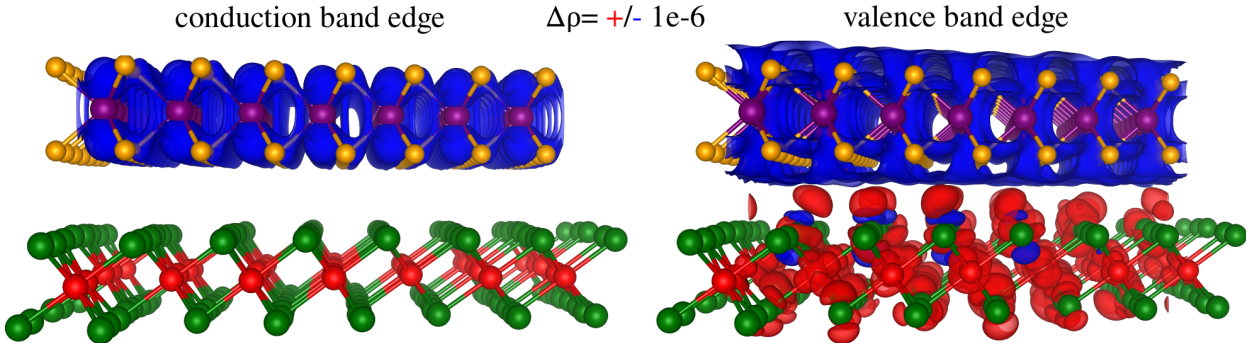

FIG. S12. Calculated spin polarization, $\Delta\rho = \rho_\uparrow - \rho_\downarrow$, for the WSe$_2$/CrI$_3$ heterostructure with a twist angle of 30°. The color red (blue) corresponds to $\Delta\rho > 0$ ($\Delta\rho < 0$). The isosurfaces correspond to isovalues (units Å$^{-3}$) as indicated. Left (right) figure takes into account WSe$_2$ conduction (valence) band edge states near the $K$ point within an energy window of 10 meV, similar as in Fig. S6.

## III. TRANSVERSE ELECTRIC FIELD — MODEL PARAMETERS

TABLE S2. Fit parameters of Hamiltonian $\mathcal{H}$ for the WSe$_2$/CrI$_3$ heterostructure for different twist angles and electric fields. We summarize the Fermi velocity $v_F$, the orbital gap $\Delta$, proximity exchange parameters $B_c$ and $B_v$, the heterostructure band gap $\Delta E$, and the calculated dipole.

| $\vartheta$ [°] | el. field [V/nm] | $v_F$ [$10^5$ m/s] | $\Delta$ [eV] | $B_c$ [meV] | $B_v$ [meV] | dipole [debye] | $\Delta E$ [eV] |
|---|---|---|---|---|---|---|---|
| 0.0000 | -2.0569 | 5.8717 | 1.4027 | -1.0052 | -0.7813 | -2.3144 | 0.5824 |
| 0.0000 | -1.0284 | 5.8732 | 1.4028 | -1.2050 | -0.8268 | -1.0980 | 0.4795 |
| 0.0000 | 0.0000 | 5.8734 | 1.4026 | -1.4598 | -0.9177 | 0.1082 | 0.3769 |
| 0.0000 | 1.0284 | 5.8754 | 1.4029 | -1.7929 | -1.0585 | 1.3224 | 0.2741 |
| 0.0000 | 2.0569 | 5.8766 | 1.4027 | -2.1939 | -1.2391 | 2.5436 | 0.1815 |
| 8.2132 | -2.0569 | 5.9539 | 1.4791 | -0.7499 | -0.0147 | -28.6859 | 0.4810 |
| 8.2132 | -1.0284 | 5.9545 | 1.4793 | -0.9326 | -0.0174 | -14.0819 | 0.3666 |
| 8.2132 | 0.0000 | 5.9536 | 1.4794 | -1.1750 | -0.0202 | 0.5475 | 0.2555 |
| 8.2132 | 1.0284 | 5.9560 | 1.4795 | -1.5086 | -0.0004 | 15.2862 | 0.1517 |
| 8.2132 | 2.0569 | 5.9550 | 1.4797 | -1.9760 | 0.0862 | 30.1340 | 0.0570 |
| 10.1583 | -2.0569 | 6.0930 | 1.6734 | -1.0064 | -0.0266 | -17.2765 | 0.3816 |
| 10.1583 | -1.0284 | 6.0930 | 1.6737 | -1.2396 | 0.0075 | -8.3129 | 0.2697 |
| 10.1583 | 0.0000 | 6.0930 | 1.6739 | -1.5534 | 0.0993 | 0.6993 | 0.1646 |
| 10.1583 | 1.0284 | 6.0906 | 1.6744 | -1.9945 | 0.3545 | 9.7808 | 0.0685 |
| 10.1583 | 2.0569 | 6.0834 | 1.6752 | -2.4441 | 1.0591 | 19.1061 | 0.0016 |
| 16.1021 | -2.0569 | 6.0856 | 1.6729 | -0.9669 | 0.3455 | -7.3068 | 0.3898 |
| 16.1021 | -1.0284 | 6.0946 | 1.6732 | -1.1888 | 0.4787 | -3.5417 | 0.2807 |
| 16.1021 | 0.0000 | 6.0839 | 1.6736 | -1.4995 | 0.7051 | 0.2474 | 0.1783 |
| 16.1021 | 1.0284 | 6.0878 | 1.6744 | -1.9277 | 1.2097 | 4.0552 | 0.0823 |
| 16.1021 | 2.0569 | 6.0535 | 1.6761 | -2.4923 | 2.6164 | 7.9234 | 0.0038 |
| 21.7868 | -2.0569 | 5.8371 | 1.4025 | -0.8998 | 0.5266 | -16.6542 | 0.5140 |
| 21.7868 | -1.0284 | 5.8388 | 1.4029 | -1.0757 | 0.6794 | -8.2206 | 0.4006 |
| 21.7868 | 0.0000 | 5.8391 | 1.4033 | -1.3179 | 0.8937 | 0.2432 | 0.2915 |
| 21.7868 | 1.0284 | 5.8373 | 1.4038 | -1.6478 | 1.2753 | 8.7598 | 0.1903 |
| 21.7868 | 2.0569 | 5.8311 | 1.4047 | -2.1360 | 1.9456 | 17.3715 | 0.0991 |
| 25.2850 | -2.0569 | 5.2185 | 1.4943 | -0.8132 | 0.7756 | -21.6021 | 0.4677 |
| 25.2850 | -1.0284 | 5.2179 | 1.4947 | -1.0148 | 1.0147 | -10.5991 | 0.3556 |
| 25.2850 | 0.0000 | 5.2156 | 1.4952 | -1.2899 | 1.4146 | 0.4402 | 0.2479 |
| 25.2850 | 1.0284 | 5.2084 | 1.4962 | -1.6829 | 2.1598 | 11.5874 | 0.1496 |
| 25.2850 | 2.0569 | 5.1822 | 1.4982 | -2.2605 | 3.8226 | 22.8533 | 0.0643 |
| 30.0000 | -2.0569 | 5.9478 | 1.4757 | -0.7577 | 0.8318 | -28.7539 | 0.4772 |
| 30.0000 | -1.0284 | 5.9476 | 1.4761 | -0.9561 | 1.0965 | -14.1298 | 0.3588 |
| 30.0000 | 0.0000 | 5.9404 | 1.4767 | -1.2263 | 1.5250 | 0.5360 | 0.2502 |
| 30.0000 | 1.0284 | 5.9343 | 1.4778 | -1.6133 | 2.3296 | 15.3416 | 0.1503 |
| 30.0000 | 2.0569 | 5.9054 | 1.4799 | -2.1768 | 4.0659 | 30.2850 | 0.0630 |

## IV.  OTHER TWIST ANGLES AND STACKINGS

Since our heterostructures have $C_3$ symmetry, and we have investigated the twist angle range between $0°$ and $30°$, we also want to address other twist angles. In particular, what happens between $30°$ and $60°$? Is the twist-angle evolution of the band edge splittings from $30°$ to $60°$ just a mirror reflection of the evolution from $0°$ to $30°$? In other words, is the splitting at $\vartheta \in [0°, 30°]$ the same as for $60° - \vartheta$? If the backfolding of the TMDC $K$ point into the $CrI_3$ Brillouin zone is the key ingredient that dictates the sign of the band edge splittings, then no surprises should occur. The only thing that changes are the magnitudes, since a geometry with $\vartheta$ has a different atomic arrangement (stacking) as a geometry with $60° - \vartheta$.

Most of our geometries contain a lot of atoms, and it is computationally very demanding to precisely calculate the electronic structure. Therefore, we focus only on $WSe_2/CrI_3$ structures with less than 150 atoms in the supercell. In particular, we look at the geometries summarized in Fig. S13 and Fig. S14, corresponding to the twist angles of $0°$, $10.2°$, $16.1°$, $21.8°$, $38.2°$, $43.9°$, $49.8°$, and $60°$, which are partner angles $\vartheta$ and $60° - \vartheta$. Before we calculate the electronic structures, we have relaxed the atomic positions, as in the main text.

For the $0°$ structure, we find that both band band edge splittings are negative, hence the negative proximity exchange parameters $B_c$ and $B_v$, see Fig. S13. Considering a different atomic arrangement, $0°+S1$, does already change the sign of the valence band splitting, while the conduction band one and the magnitudes remain the same. However, the $0°+S1$ structure is energetically less favorable by about 12.6 meV. Apparently, since there are only 20 atoms in the supercell geometry, the precise atomic stacking strongly influences the valence band edge splitting. Additionally, the Cr and I atoms have opposite magnetic moments, providing opposite local exchange fields. Looking at the $0°+S2$ structure, both band edge splittings are again negative but diminished in magnitude. Also the $0°+S2$ structure is energetically less favorable by about 83.6 meV. For the $60°$, $60°+S1$, and $60°+S2$ structure, we find them to be 83.7, 12.6, and 0.1 meV higher in energy than the $0°$ one. From this energetic analysis, and comparing the extracted proximity exchange parameters, we see that each $0°$ structure has a $60°$ partner.

Moreover, we look at a geometry with $43.9°$, which is the partner angle of $16.1°$. The $43.9°$ structure is 23.2 meV lower in energy than the $16.1°$ one, but proximity exchange couplings are similar, see Fig. S13. In addition, the $16.1°+S1$ ($43.9°+S1$) structure is 1.8 meV (6.8 meV) higher (lower) in energy, while exchange parameters barely change. Finally, see Fig. S14, the $10.2°+S1$ and $49.8°$ structures are 8.2 meV and 25.2 meV lower in energy compared to the $10.2°$ structure. Proximity exchange couplings are again barely affected when comparing the shifted structure and the partner twist angle structure. The $21.8°+S1$ and $38.2°$ structures are 28.7 meV higher and 157.3 meV lower in energy compared to the $21.8°$ structure. Proximity exchange couplings are again similar in their values.

This shows that twisting from $30°$ to $60°$ leads to similar results — apart from differences in magnitude owed to the stacking — as twisting from $30°$ to $0°$. Moreover, we find that in larger supercells the atomic stacking only marginally influences proximity exchange couplings, while in small supercells the precise stacking matters a lot, similar as in Ref. [1]. We believe that an averaging effect takes place in larger supercell geometries, where a lot of atoms are involved in forming the TMDC band edge states, which can locally pick up different exchange fields, but globally lead to the same spin splitting.

Since there is already a difference in the valence band splitting for the $0°$ and $0°+S1$ structures, we further analyze the latter to find out about the origin of this. Similar as above, we have calculated the projected band structure, DOS, and TMDC band edge spin polarizations for the $0°+S1$ structure, see Figs. S15, S17, and S16. Comparing the projected band structures and DOS, we cannot identify a significant difference that would lead to the reversal of the valence band splitting. From the spin polarization, we also find no difference for the conduction band edge, but the valence band edge is markedly different, compare Fig. S6 and Fig. S16. In order to resolve this, in Fig. S18 we explicitly compare the spin resolved atomic DOS contributions of $0°$ and $0°+S1$ structures. Apart from the slightly different band alignment — which can be seen best in the W and Se comparison figure — significant differences arise in the energy window from $-1.5$ to $-1$ eV below the Fermi level (grey disks). In particular, the contribution of I atoms and spin up contribution of W atoms is markedly modified. This confirms, that there is a delicate balance between the couplings of TMDC states to the spin up and spin down manifolds of $CrI_3$, which is strongly stacking dependent and can lead to a reversed valence band splitting.

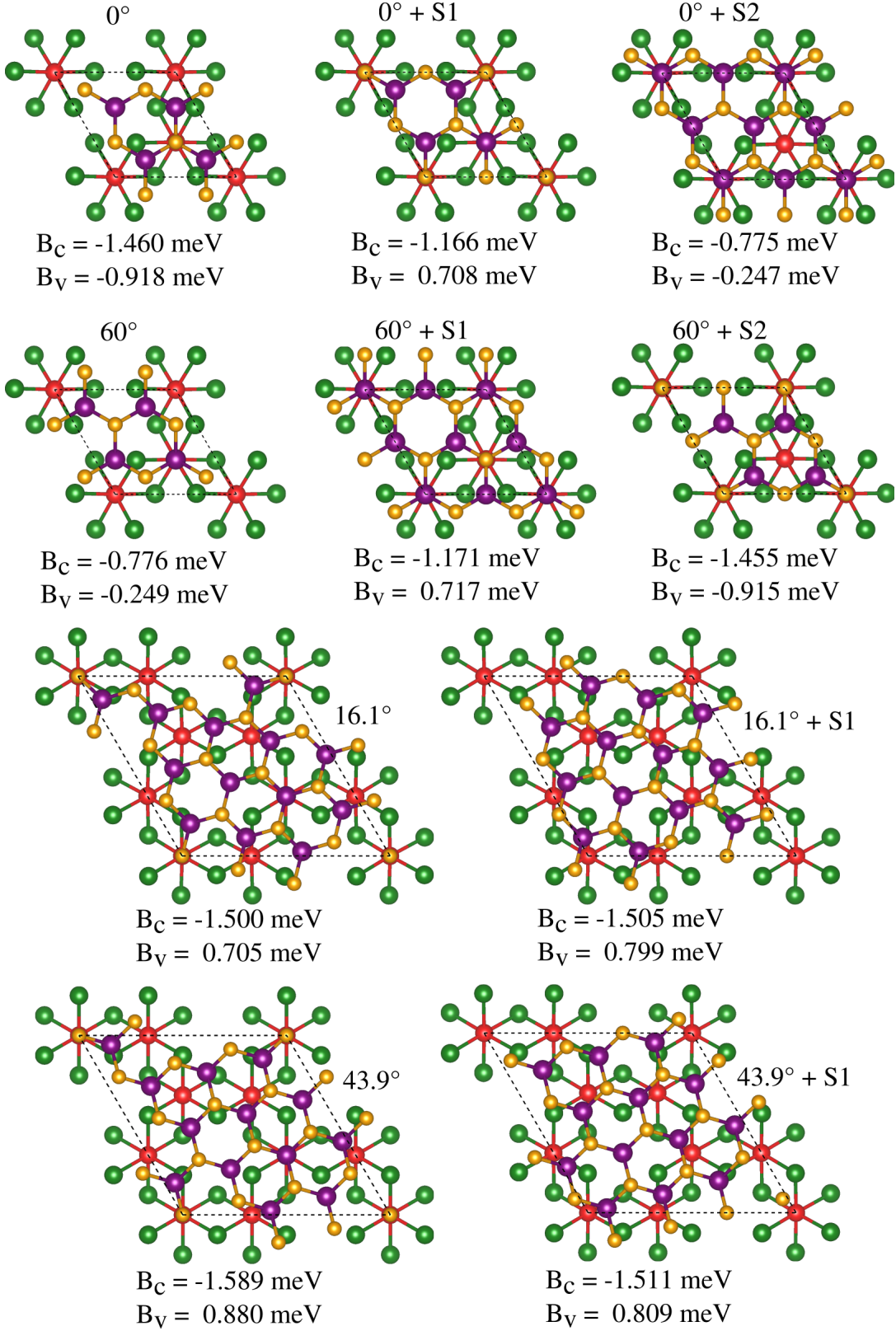

FIG. S13. WSe$_2$/CrI$_3$ geometries corresponding to twist angles of 0°, 16.1°, 43.9°, and 60°. For each twist angle we consider different atomic arrangements, i.e., we shift the TMDC layer with respect to the CrI$_3$. For each structure we label the twist angle (+ shift) and the corresponding proximity exchange parameters $B_c$ and $B_v$. The 0° and 16.1° structures are the ones from the main text. Cr = red, I = green, W = purple, and Se = orange.

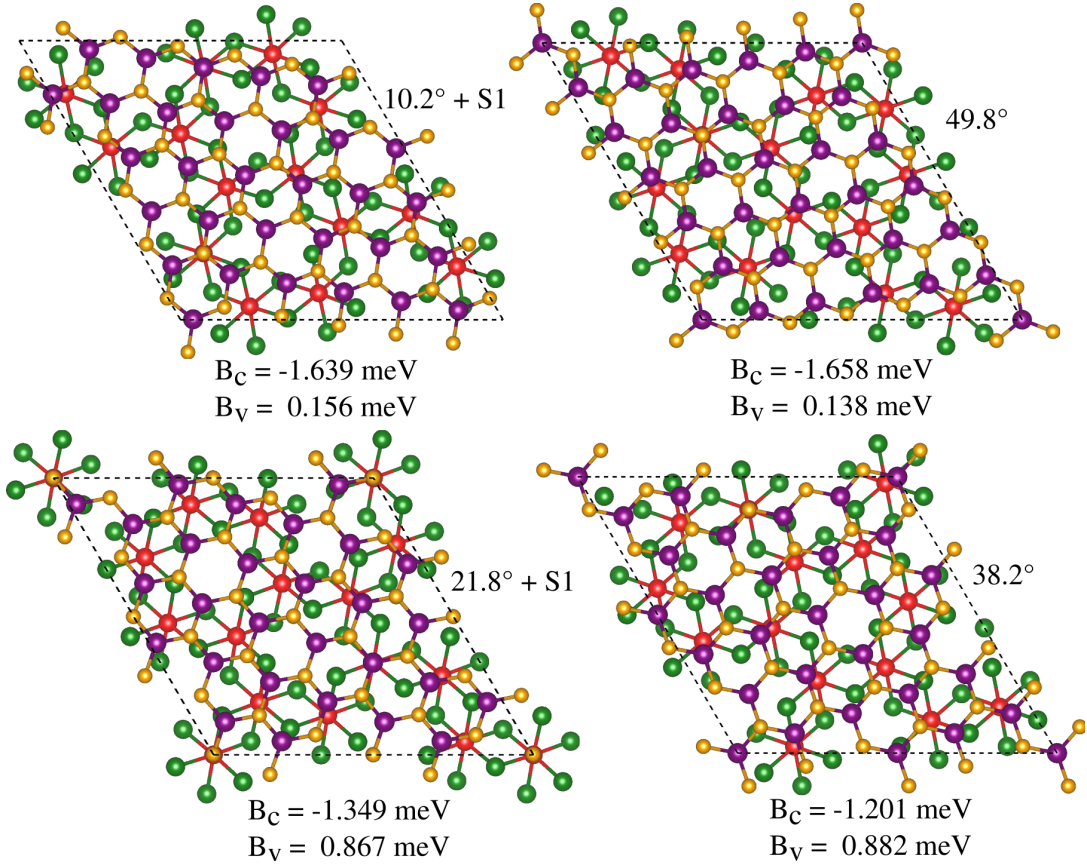

FIG. S14. WSe$_2$/CrI$_3$ geometries corresponding to twist angles of 10.2°, 21.8°, 38.2°, and 49.8°. For each structure we label the twist angle (+ shift) and the corresponding proximity exchange parameters $B_c$ and $B_v$. Cr = red, I = green, W = purple, and Se = orange.

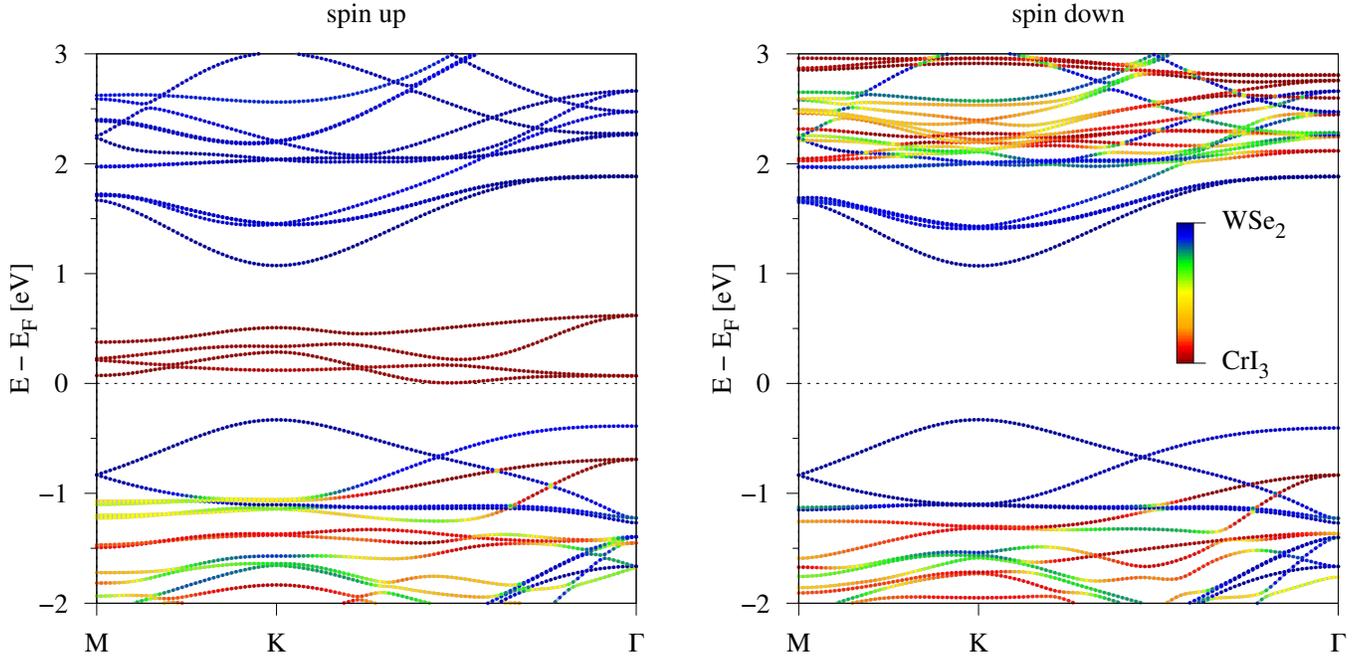

FIG. S15. Same as Fig. S3, but for the 0°+S1 structure.

**conduction band edge**          **valence band edge**

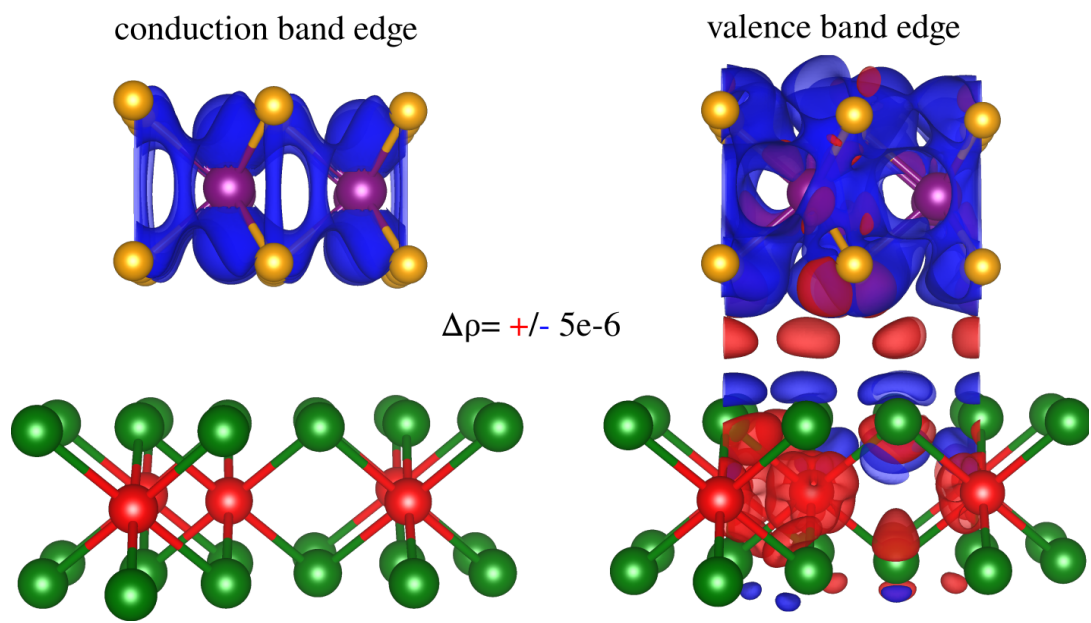

$\Delta\rho = +/- \ 5e\text{-}6$

FIG. S16. Calculated spin polarization, as in Fig. S6, but for the 0°+S1 structure.

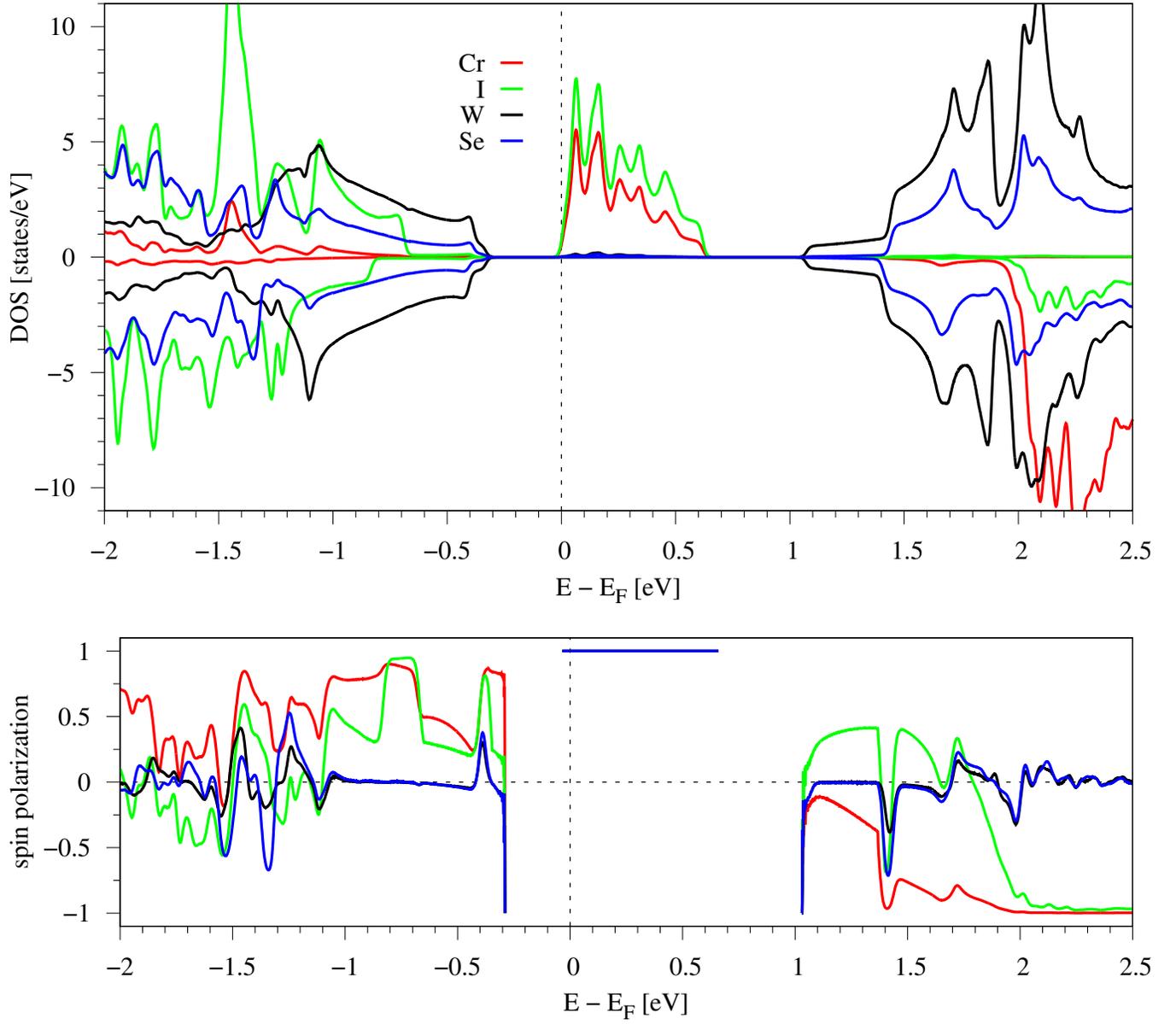

FIG. S17. Same as Fig. S4, but for the 0°+S1 structure.

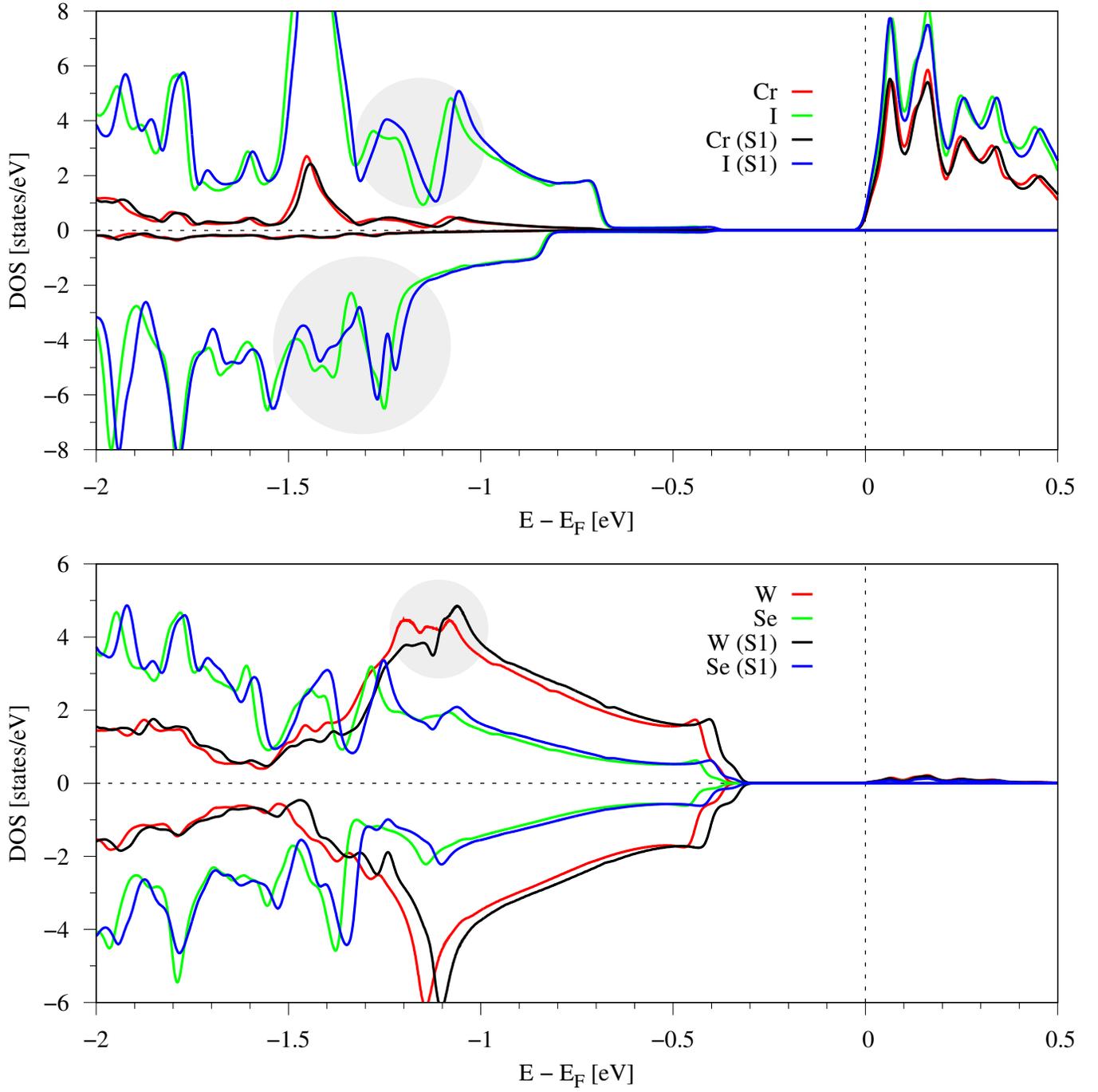

FIG. S18. Comparison of the spin resolved atomic DOS between 0° and 0°+S1 structures. In the top (bottom) figure, we compare the Cr and I (W and Se) contributions of the WSe$_2$/CrI$_3$ heterostructure. Grey disks highlight pronounced differences.

## V. MAPPING OF LOCAL STACKINGS

From the experimental point of view, where no or only small strain is present, the heterostructures will be noticeably different. For example, for the 0° structure, we have employed strains of about ±2% for the individual layers to make them commensurate. Therefore, we end up with a computationally feasible scenario of 20 atoms in the supercell, at the cost of relatively large strain. When we want to reduce strain, in order to more realistically reproduce experimental conditions, we will end up with much more atoms in the supercell. For example, in Fig. S19, we show an experimentally more realistic 0° WSe$_2$/CrI$_3$ heterostructure with 2123 atoms and nearly unstrained monolayers. To calculate the electronic structure would be computationally very demanding and is beyond our limits. However, as we show in Fig. S19, the structure consists of different local stackings from our smaller strained 0° supercell that we have already calculated. From our previous considerations for the 0° structure, we know that the proximity exchange couplings strongly depend on the stacking configuration. In other words, experimentally the proximity exchange will vary locally in space across the heterostructure. To fully map out the 'unstrained' heterostructure would still require a lot of computational efforts, since each local stacking needs to be calculated. Fig. S19 essentially serves as a proof of principle for the above discussion.

For the other twist angles, we have not seen such a drastic stacking dependence, since the supercells were already large enough such that an averaging effect takes place. Still, proximity exchange will vary locally in experimental setups.

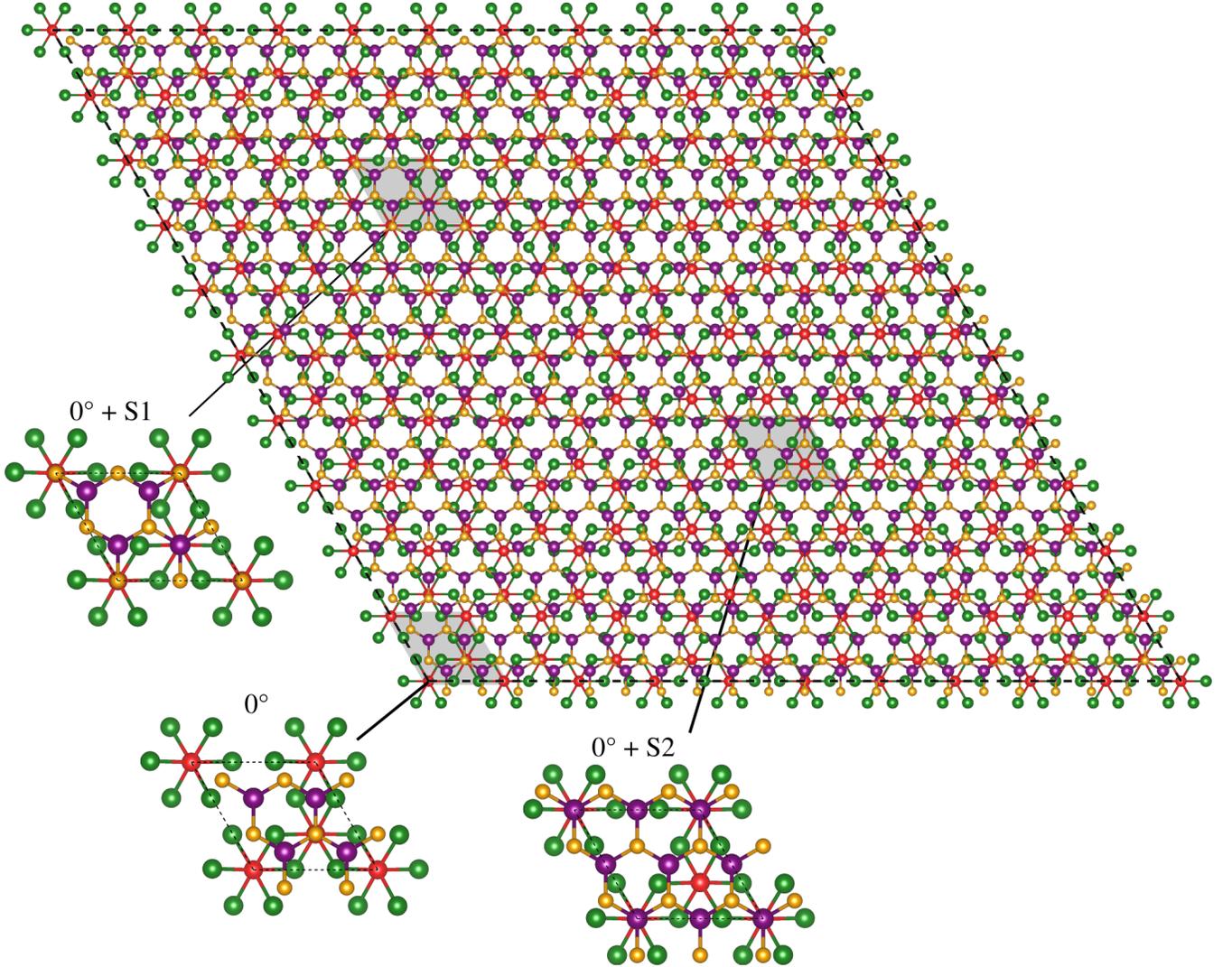

FIG. S19. An experimentally *more realistic* 0° WSe$_2$/CrI$_3$ heterostructure (2123 atoms), where we consider a 21 × 21 supercell of WSe$_2$ on a 10 × 10 supercell of CrI$_3$, corresponding to strains of 0% and 0.5%. The different stackings from our 0° *strained* heterostructure from Fig. S13 are locally indicated.

# VI. INFLUENCE OF THE STRAIN DISTRIBUTION

As already mentioned in the main text, monolayers of TMDCs and CrI$_3$ are based on hexagonal unit cells, with lattice constants of $a = 3.288$ Å (MoSe$_2$), $a = 3.282$ Å (WSe$_2$), and $a = 6.867$ Å (CrI$_3$) [5–7], which need to be strained in the twisted heterostructures, in order to form commensurate supercells for periodic DFT calculations. We already know that biaxial strain strongly influences the band gap, spin-orbit splittings and spin-valley properties of monolayer TMDCs [2, 8], as well as the band gap of CrI$_3$ [9]. Therefore we keep the strain as small as possible and nearly equally distribute it among the monolayers, see Table I. However, experimentally such a strain distribution among the layers will not arise. Still, for some of the supercells, in particular 10.2° and 16.1°, the strain is very small and we can nearly view them as in an experimental setup.

What is the impact of the strain distribution on the proximity exchange? To answer this question, we focus again on the WSe$_2$/CrI$_3$ heterostructure with a twist angle of 0°. By tuning the in-plane lattice constant, the total strain cannot be reduced, as commensurability needs to be fullfiled, but the strain distribution can be changed. Again, we perform structural relaxations before calculating the electronic properties for the different strain distributions.

In Table S3, we summarize the relevant results. We find that all extracted parameters depend on the strain distribution. As already known, the band gap $\Delta$ of the WSe$_2$ can be strongly tuned by strain, see also Ref. [2]. Most important for us is the tunability of proximity exchange parameters $B_c$ and $B_v$. We find that, within our investigated strain distribution limits, the two parameters can be changed by at maximum $\pm 20\%$. In other words, our calculated results are representative for experimental comparison, since the values change with strain, but the order of magnitude remains. The reason for this tunability of proximity exchange is most likely related to the tunability of the CrI$_3$ magnetism with strain, see Table S4. One can expect a similar strain dependence for the other investigated supercells, corresponding to the different twist angles.

It is also worth to notice, that the CrI$_3$ in-gap states bandwidth $\Delta W$ depends linearly on the lattice constant. In contrast, the heterostructure bandgap $\Delta E$ does not, since the valence band edge maximum of WSe$_2$ performs a $K \rightarrow \Gamma$ transition upon positive strain larger than $\sim 3\%$ [2].

TABLE S3. Structural information and fit parameters of Hamiltonian $\mathcal{H}$ for the WSe$_2$/CrI$_3$ heterostructures with a twist angle of 0° and different strain distributions. We summarize the biaxial strains $\varepsilon$ applied to WSe$_2$ and CrI$_3$, the calculated dipoles, the relaxed interlayer distances $d_{\text{int}}$, the Fermi velocity $v_F$, the orbital gap $\Delta$, proximity exchange parameters $B_c$ and $B_v$, the heterostructure band gap $\Delta E$, and the CrI$_3$ in-gap states bandwidth $\Delta W$.

| $\varepsilon_{\text{WSe}_2}$ [%] | $\varepsilon_{\text{CrI3}}$ [%] | dipole [debye] | $d_{\text{int}}$ [Å] | $v_F$ [$10^5$ m/s] | $\Delta$ [eV] | $B_c$ [meV] | $B_v$ [meV] | $\Delta E$ [eV] | $\Delta W$ [eV] |
|---|---|---|---|---|---|---|---|---|---|
| 4.4204 | -0.0046 | 0.1268 | 3.5312 | 5.5502 | 1.1853 | -1.5814 | -0.9870 | 0.2835 | 0.5009 |
| 3.3165 | -1.0617 | 0.1169 | 3.5205 | 5.7209 | 1.2878 | -1.4937 | -0.9403 | 0.3502 | 0.5496 |
| 2.2126 | -2.1188 | 0.1082 | 3.5000 | 5.8734 | 1.4026 | -1.4598 | -0.9177 | 0.3769 | 0.6002 |
| 1.1087 | -3.1759 | 0.0960 | 3.4790 | 6.0018 | 1.5311 | -1.4020 | -0.8858 | 0.3536 | 0.6542 |
| 0.0048 | -4.2331 | 0.0740 | 3.5129 | 6.0577 | 1.6717 | -1.1719 | -0.7814 | 0.3207 | 0.7181 |

TABLE S4. Calculated averaged atomic magnetic moments in units of $\mu_B$ of the WSe$_2$/CrI$_3$ heterostructure with a twist angle of 0° and different strain distributions. For Se, we consider only the atoms at the interface.

| $\varepsilon_{\text{WSe}_2}$ [%] | $\varepsilon_{\text{CrI3}}$ [%] | W | Se | Cr | I |
|---|---|---|---|---|---|
| 4.4204 | -0.0046 | -0.0029 | -0.0037 | 3.3904 | -0.1573 |
| 3.3165 | -1.0617 | -0.0027 | -0.0035 | 3.3710 | -0.1533 |
| 2.2126 | -2.1188 | -0.0026 | -0.0034 | 3.3511 | -0.1491 |
| 1.1087 | -3.1759 | -0.0025 | -0.0033 | 3.3299 | -0.1448 |
| 0.0048 | -4.2331 | -0.0021 | -0.0028 | 3.3149 | -0.1422 |

## VII. CALCULATED ABSORPTION SPECTRA

The calculated absorption spectra are presented in Fig. S20 for the different systems (WSe$_2$/CrI$_3$ and MoSe$_2$/CrI$_3$) and assuming different top dielectric environments (vacuum or hBN), as discussed in the main text. The calculations follow the exciton parameters of our previous study, discussed in detail in the Supplemental Material of Ref. [10], and the model Hamiltonian parameters given in Table S2. We note that the dependence of the first absorption peak is mainly controlled by the value of $\Delta$ (almost the bang gap of the material). For the twist angles 10.2° and 16.1°, the values of $\Delta$ are $\sim 0.2$ eV larger than the others (see Table S2), which is related to the TMDC lattice constant within the twisted heterostructures, and thus the absorption peaks are blue-shifted.

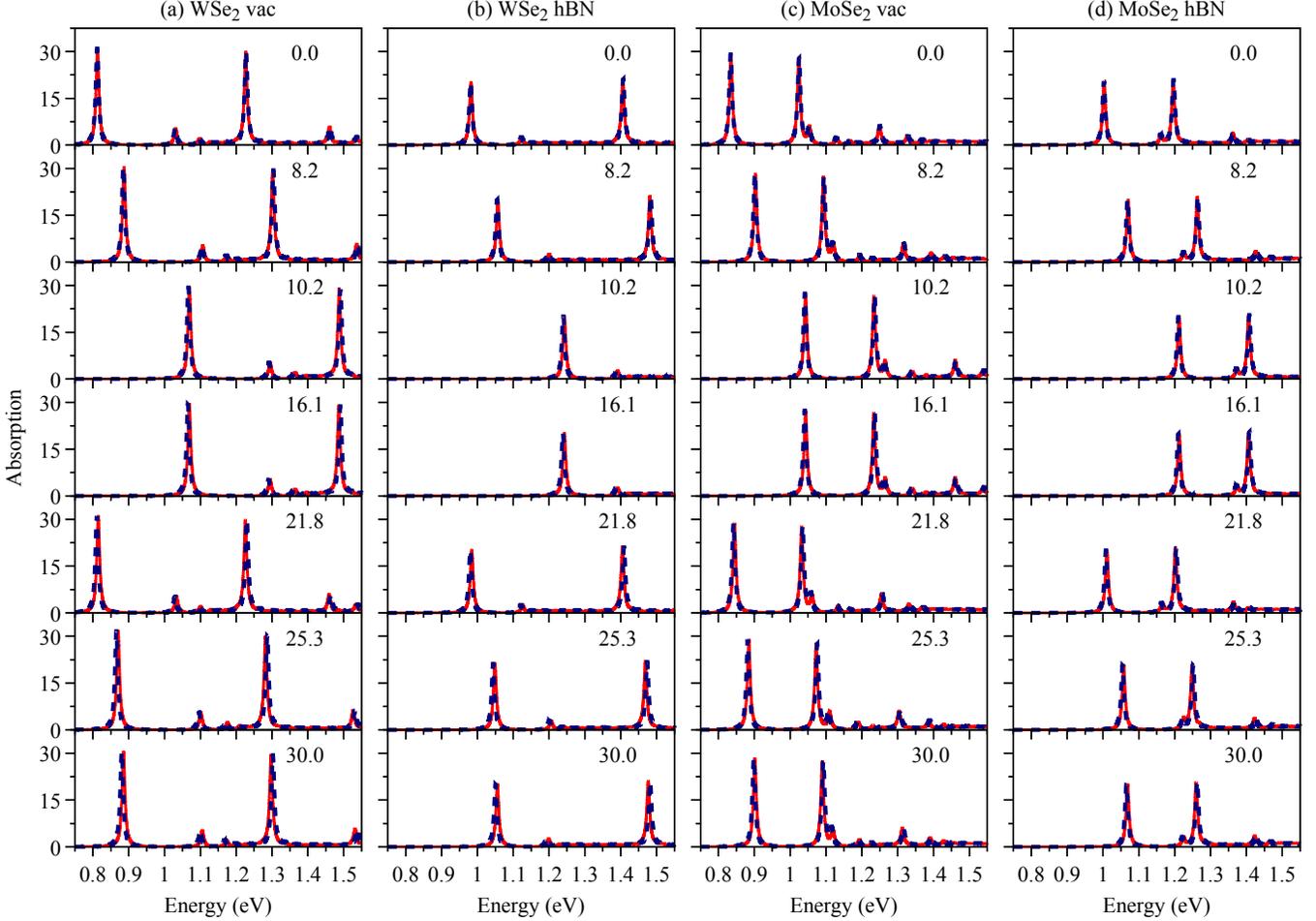

FIG. S20. Representative absorption spectra of intralayer excitons for (a) vacuum/WSe$_2$/CrI$_3$, (b) hBN/WSe$_2$/CrI$_3$, (c) vacuum/MoSe$_2$/CrI$_3$ and (d) hBN/MoSe$_2$/CrI$_3$ for the different twist angles considered in our study at zero electric field.

\clearpage

\end{document}